# Efficient production of large-size optical Schrödinger cat states

Evgeny V. Mikheev[1], Alexander S. Pugin[1,2], Dmitriy A. Kuts[1], Sergey A. Podoshvedov[1] and Nguyen Ba An[3,4]

[1] *Laboratory of Quantum Information Processing and Quantum Computing, Institute of Natural and Exact Sciences, South Ural State University (SUSU), Lenin Av. 76, Chelyabinsk, Russia*
Email: sapodo68@gmail.com
[2] *Department of Applied Mathematics and Programming, Institute of Natural and Exact Sciences, South Ural State University (SUSU), Lenin Av. 76, Chelyabinsk, Russia*
[3] *Thang Long Institute of Mathematics and Applied Sciences (TIMAS), Thang Long University, Nghiem Xuan Yem, Hoang Mai, Hanoi, Vietnam*
[4] *Institute of Physics, Vietnam Academy of Science and Technology (VAST), 18 Hoang Quoc Viet, Cau Giay, Hanoi, Vietnam.*
E-mail: nban@iop.vast.ac.vn

**Abstract.** We present novel theory of effective realization of large-size optical Schrödinger cat states, which play an important role for quantum communication and quantum computation in the optical domain using laser sources. The treatment is based on the $\alpha$-representation in infinite Hilbert space which is the decomposition of an arbitrary quantum state in terms of displaced number states characterized by the displacement amplitude $\alpha$. We find analytical form of the $\alpha$-representation for both even and odd Schrödinger cat states which is the key for proposing their generation schemes. Two schemes are proposed for generating even/odd Schrödinger cat states of large size $|\beta|$ ($|\beta| \geq 2$) with high fidelity $F$ ($F \approx 0.99$). One scheme relies on an initially offline prepared two-mode entangled state with a fixed total photon number, while the other scheme uses separable photon Fock states as the input. In both schemes, the desired states are conditionally generated if the corresponding measurement outcomes are fixed. Conditions for obtaining states useful for quantum information processing are established and success probabilities for their generation are evaluated.

It is known that a potentially quantum computer can effectively implement intractable algorithms such as large integer factoring[1] and unsorted data search[2] which cannot be implemented by computers operating under classical laws. But realization of the quantum computer requires effective performance of a universal set of deterministic gate operations over a large set of qubits[3]. Also, qubits are exposed to influence of the environment, requiring good fault-tolerant computational systems. All these impose highly stringent requirements on the physical system where qubits and quantum gates are realized. Different physical systems might be used to implement different quantum protocols. In particular, as light has the maximally possible speed of propagation and weakly interacts with the surrounding noisy environment, optical systems are put in one row with atomic ones in the design of possible configurations of the quantum computer.

Although there are many proposed approaches for optical quantum computers, none of them are completely satisfactory since they are quite complex and/or restricted in application. For example, realization of deterministic gate operations[4] would require an unacceptably huge number of additional operations[5,6]. So, one can hardly say that the issue of optical quantum information processing (QIP) has been finally resolved[7] and the question of how to efficiently exploit the optical resources (interaction mechanisms, approaches, suitable states) for QIP remains of great interest. Up to now, three approaches for optical QIP are developed within the discrete-variable (DV)[5], continuous-variable (CV)[8] and combined discrete-continuous-variable (DV-CV) frameworks[9]. These approaches use one of aspects of the particle-wave duality[7] or both of them[10,11]. Each approach has its own inherent advantages and drawbacks. Namely, the DV approach uses photons that interact very weakly with each other so two-qubit operations can be realized only in a non-deterministic manner[12]. Instead, quantum protocols with CV states can be implemented deterministically, but the fidelity is limited due to the fact that CV entangled states such as two-mode squeezed vacuum state does not carry maximum entanglement[13]. Commonly used optical states are the so-called optical Schrödinger cat states (SCSs) $a_0|-\alpha\rangle \pm a_1|\alpha\rangle$, with $|\pm\alpha\rangle$ being coherent states with macroscopic continuous amplitudes $\pm\alpha$ and $a_0, a_1$ normalization coefficients. These states can also be referred to as quantum superpositions of two out-of-phase light pulses. The size of coherent components $|\alpha|$ is of crucial importance in the experiments to test quantum foundations and quantum information technologies[14-18]. Generally $|-\alpha\rangle$ and $|\alpha\rangle$ are not strictly orthogonal to each other. But, since their overlap is determined by $|\langle\alpha|-\alpha\rangle| = exp(-2|\alpha|^2)$, for $|\alpha| \geq 2$ one has $|\langle\alpha|-\alpha\rangle| \leq 3 \cdot 10^{-4} \approx 0$, so such SCSs can be treated as good qubits. They are called large-size SCSs where "large-size" practically implies $|\alpha| \geq 2$. However, it is very difficult to produce such large-size SCSs in realistic conditions with existing third-order nonlinearities. Although a lot of progress has been made over last time[19-26], size of the generated SCSs as well as their low generation rate still leave much to be improved for desired practical protocols. In other words, the realization of sufficiently large-size SCSs remains questionable and is worth further tremendous efforts. In this connection, the DV-CV approach with the so-called hybrid states turns out to be a promising direction since the combination of two different physical systems could provide new capabilities to more efficiently implement optical quantum protocols[27-36].

Since the direct implementation of the SCSs[37] is currently impossible due to the smallness of the third-order optical nonlinearity, it makes sense to consider other methods[38] that can approximate the output of the $\chi^{(3)}$ nonlinearity with high fidelity. A scheme for generating SCSs by feeding a squeezed vacuum into beam splitter and counting photons in auxiliary mode was considered in[39]. It was also shown[40] that any single-mode quantum state can be generated from the vacuum by alternate applications of displacement operations combined with single photons. Recently, the techniques of photon subtraction and photon addition are fairly common for generating different types of SCSs[17,21,25,26,41-43]. These techniques are widely demonstrated in modern optical experiments[44-50].

Here, we present novel ways to generate even/odd SCSs of large size that could be directly used in work of quantum computer. The method is based on introduction of the so-called $\alpha$-representation of even/odd SCSs which is their decomposition in base of the displaced number states[17]. One method is based on the pre-generation of a two-mode entangled state with a fixed number $n$ of photons[50]. Photon subtraction from the displaced number state[51,52] of the original entangled one in auxiliary mode allows one to generate the states that under certain conditions approximate either even or odd large-size SCSs with fidelity higher than 0.99 that are suitable for quantum protocols. This approach allows one to find strategy for generating auxiliary two-mode entangled states taking into account experimental conditions and imperfections imposed in reality. The method could be considered ideal, provided that the auxiliary two-mode entangled states[50] is already prepared in advance. In order to avoid the efforts associated with the generation of a rather complicated entangled state, we also develop another method for conditional generation of even/odd large-size SCSs by mixing photon

Fock states on beam splitters followed by displacing the auxiliary modes and subsequent measuring their photon numbers by photo-detectors. Both the schemes allow us to effectively generate even/odd SCSs with large size.

Our paper is structured as follows. We begin with a general description approach of Schrödinger cat states in infinite Hilbert space. To do this, we introduce the concept of $\alpha$-representation of Schrödinger cat states. This gives us the opportunity to introduce in Section "Schrödinger cat states" the concept of Schrödinger cat qudits (SCQs) being excellent approximation of the Schrödinger cat states. In next section, we consider the idea of generation of even/odd Schrödinger cat qudits by using the offline created two-mode entangled state containing $n$ photons in total. We show that extracting the displaced photon state from the given quantum channel allows us to observe the generation of both even and odd SCQs with size $\beta = 2$. Graphs of success probabilities are presented. Experimental parameters at which the generation of two-mode entangled quantum channel is possible, are collected in Tables. Section "Generation of SCQ from Fock states" introduces a new idea of the probabilistic generation of even and odd Schrödinger cat qudits of large size from initial number states as tensor product. A system of beam splitters and displacement operators to make the original photons unitarily evolve in the desired manner is used. The final registration of vacuum states in auxiliary modes allows us to conditionally generate even/odd SCQs of large size. The corresponding values of the parameters at which this generation is possible are presented in the tables. The Wigner functions of the generated and ideal states confirming the correctness of the developed method are plotted. In the final section 5, we summarize the resulting material and argue about the effectiveness of the proposed approaches for generating SSQs of large size. Section Methods involves two subsections one of the concerns derivation of $\alpha$-representation of Schrödinger cat states, while other derives a main formula appearing in Section "Generation of SCQ from Fock states".

## Results

**Schrödinger cat qudits.** The even/odd SCSs $|\beta_\pm\rangle$ with size $|\beta|$ are defined by

$$|\beta_+\rangle = N_+(|-\beta\rangle + |\beta\rangle), \quad (1)$$

$$|\beta_-\rangle = N_-(|-\beta\rangle - |\beta\rangle), \quad (2)$$

where $N_\pm = \left(2(1 \pm exp(-2|\beta|^2))\right)^{-1/2}$ are the normalization factors, which in general depend on $|\beta|$, and the notations $|\pm\beta\rangle$ mean coherent states with amplitudes $\pm\beta$. The amplitude $\beta$ is generally complex, but here and in the following, for simplicity, it is assumed to be real and positive (i.e., $\beta > 0$). Then the amplitude $\beta$ of the SCS is regarded as its size. It is also called superposition of coherent states. We are going to use both designations of the state under study.

The even/odd SCSs are obviously orthogonal to each other, $\langle\beta_-|\beta_+\rangle = 0$, as the photon numbers in $|\beta_+\rangle$ ($|\beta_-\rangle$) are even (odd). In this paper we are working with the so-called $\alpha$-representation. The $\alpha$-representation of any state is determined in infinite Hilbert space of the displaced number states (47) characterized by the displacement amplitude $\alpha$[17,18]. Precisely, the $\alpha$-representation of an arbitrary state is its decomposition over the basis states $\{|k, \alpha\rangle; k = 0,1, \dots, \infty\}$ of the displaced number states. In the case of the optical SCSs, we have from Eq. (57) in Section Methods:

$$|\beta_+\rangle = N_+ exp\left(-\frac{|\alpha|^2+|\beta|^2}{2}\right) \sum_{k=0}^{\infty} a_k^{(+)} |k, \alpha\rangle, \quad (3)$$

$$|\beta_-\rangle = N_- exp\left(-\frac{|\alpha|^2+|\beta|^2}{2}\right) \sum_{k=0}^{\infty} a_k^{(-)} |k, \alpha\rangle, \quad (4)$$

where the decomposition coefficients $a_k^{(\pm)}$ for arbitrary $\alpha$ and $\beta$ are given by

$$a_k^{(\pm)} = \frac{1}{\sqrt{k!}}\left(e^{-\alpha^*\beta}(-\alpha - \beta)^k \pm e^{\alpha^*\beta}(-\alpha + \beta)^k\right). \quad (5)$$

It is possible to directly check that the normalization condition is satisfied for both even and odd SCSs, i.e., $N_\pm^2 exp(-(|\alpha|^2 + |\beta|^2)) \sum_{n=0}^{\infty} |a_n^{(\pm)}|^2 = 1$ hold for any values of the parameters $\alpha$ and $\beta$.

It is well known that the size of SCSs generated by direct use of $\chi^{(3)}$ nonlinearities[14] cannot be large enough due to the tininess of $\chi^{(3)}$ nonlinearities available in all existing nonlinear crystals. As will be seen later, the $\alpha$-representations (3, 4) prove to be useful for the problem of generating sufficiently large-size optical Schrödinger cats. Quantum engineering allows the replacement of the original infinite CV state with its finite version which represents a truncated superposition of just $n + 1$ terms in the corresponding $\alpha$-representation, with $n$ being some integer. That is, we can approximate the SCSs (3, 4) by the following states

$$|\Psi_n^{(\pm)}\rangle = N_n^{(\pm)} \sum_{k=0}^{n} b_k^{(\pm)} |k, \alpha\rangle, \qquad (6)$$

with $b_k^{(\pm)}$ some expansion coefficients to be specified later and $N_n^{(\pm)} = \left(\sum_{k=0}^{n} |b_k^{(\pm)}|^2\right)^{-1/2}$ the normalization factors. We can also speak about replacing optical original SCSs in Eqs. (3, 4), which are CV states residing in an infinite Hilbert space, by the states in Eq. (6), which are DV states residing in a finite Hilbert space of dimension $d = n + 1$. The degree of validity for such a replacement can be assessed by the fidelity $F_n^{(\pm)} = tr\left(\varrho_n^{(\pm)} \varrho^{(SCS)}\right)$, with $tr$ denoting the trace over the state in parentheses, $\varrho^{(SCS)}$ the density matrix of the original pure states (3, 4) and $\varrho_n^{(\pm)}$ the density matrix of the states (6). The fidelity value lies in the range from 0 up to 1. If the fidelity is equal to 1, then the compared states are identical to each other. Conversely, if the fidelity is equal to 0, then such states are orthogonal to each other. The bigger value the fidelity acquires the closer to each other are the two compared states. In the case of the optical SCSs (3, 4) and their truncated versions (6), the fidelity can be written as

$$F_n^{(+)} = \left|\langle\beta_+|\Psi_n^{(+)}\rangle\right|^2 = N_+^2 N_n^{(+)2} exp(-\mathbb{a}^2) \left|\sum_{k=0}^{n} a_k^{(+)*} b_k^{(+)}\right|^2, \qquad (7)$$

$$F_n^{(-)} = \left|\langle\beta_-|\Psi_n^{(-)}\rangle\right|^2 = N_-^2 N_n^{(-)2} exp(-\mathbb{a}^2) \left|\sum_{k=0}^{n} a_k^{(-)*} b_k^{(-)}\right|^2, \qquad (8)$$

where $\mathbb{a} = \sqrt{|\alpha|^2 + |\beta|^2}$.

Here we confine ourselves to considering only the purely imaginary amplitude of the displacement of the basis states, i.e, the displacement amplitude is assumed to be equal to $i\alpha$, with $\alpha$ a real number. The choice of such purely imaginary displacement amplitude $i\alpha$ looks convenient since then the value of the parameter $i\alpha$ will lie symmetrically with respect to the quantities $-\beta$ and $\beta$ on the phase plane. It allows us to rewrite the decomposition coefficients (5) of the $\alpha$-representation of SCSs in more compact forms as (59, 60)

$$a_k^{(+)} = \frac{2(i\mathbb{a})^k}{\sqrt{k!}} cos(\alpha\beta + k(\varphi + \pi/2)), \qquad (9)$$

$$a_k^{(-)} = \frac{2(i\mathbb{a})^k}{\sqrt{k!}} sin(\alpha\beta + k(\varphi + \pi/2)), \qquad (10)$$

where the relative phase is $\varphi = arctang(\alpha/\beta)$. We can see that the coefficients may not be equal to zero for arbitrary values of $k$. We obtain standard form of the coefficients of the even/odd SCSs (49, 50) in the Fock or number state basis (or, the same, in the 0-representation) if we take $\alpha = 0$ in Eqs. (9, 10). The division into 'even' and 'odd' takes place exclusively in the 0-representation of the SCSs. In any other representations (i.e., representations with $\alpha \neq 0$), this division into 'even' and 'odd' is meaningless, because, as seen from (9, 10), they contain both even and odd displaced states. Nevertheless, we adhere to the standard division of the SCSs into even and odd ones even in arbitrary $\alpha$-presentations (9, 10).

By numerical calculations we find out that the best way to approximate the original SCSs (1,2) with highest fidelity is to set the expansion coefficients in Eq. (6) to be proportional to those in Eqs. (9, 10), say, in the following way: $b_k^{(+)} = a_k^{(+)}/2$ and $b_k^{(-)} = a_k^{(-)}/2$. Let us denote the states (6) with such setting for the coefficients by $|\Psi_n^{(S+)}\rangle$ and loosely call them Schrödinger cat qudits (SCQs) of dimension $d = n + 1$ which have the following form

$$|\Psi_n^{(S+)}\rangle = N_n^{(S+)} \sum_{k=0}^{n} \left(a_k^{(+)}/2\right)|k, i\alpha\rangle =$$
$$N_n^{(S+)} \sum_{k=0}^{n} \left((i\mathbb{a})^k/\sqrt{k!}\right) \cos(\alpha\beta + k(\varphi + \pi/2))|k, i\alpha\rangle \quad (11)$$

and

$$|\Psi_n^{(S-)}\rangle = N_n^{(S+)} \sum_{k=0}^{n} \left(a_k^{(-)}/2\right)|k, i\alpha\rangle =$$
$$N_n^{(S-)} \sum_{k=0}^{n} \left((i\mathbb{a})^k/\sqrt{k!}\right) \sin(\alpha\beta + k(\varphi + \pi/2))|k, i\alpha\rangle, \quad (12)$$

with the corresponding normalization factors

$$N_n^{(S+)} = \left(\sum_{k=0}^{n} (\mathbb{a}^{2k}/k!) \cos^2(\alpha\beta + k(\varphi + \pi/2))\right)^{-1/2}, \quad (13)$$
$$N_n^{(S-)} = \left(\sum_{k=0}^{n} (\mathbb{a}^{2k}/k!) \sin^2(\alpha\beta + k(\varphi + \pi/2))\right)^{-1/2}, \quad (14)$$

by virtue of Eqs. (9,10). Then, we can derive from Eqs. (7,8) the expressions for the fidelities $F_n^{(S\pm)}$ between the original SCSs (1,2) and the approximated ones, i.e., the SCQs in Eqs. (11,12):

$$F_n^{(S\pm)} = \left(N_\pm^2 N_n^{(S\pm)2} \exp(-\mathbb{a}^2)/4\right) \sum_{k=0}^{n} \left|a_k^{(\pm)}\right|^2. \quad (15)$$

The functions $F_n^{(S\pm)}$ depend not only on $n$ but also on two independent variables $\alpha$ and $\beta$. Three-dimensional plots of $F_n^{(S+)}$ and $F_n^{(S-)}$ in dependency on $\alpha$ and $\beta$ are shown in Figs. 1 and 2, respectively, where the value of $n$ varies from 2 up to 9. A general rule is observed. If the value of $n$ increases, then the values of the fidelities $F_n^{(S\pm)}$ increase too and approaches 1 starting from some large enough value of $n$ (say, $n \geq 9$). The range of the values of $\alpha$ and $\beta$, in which high fidelities are achieved, is also increased. Visually, already with $n = 9$ the SCQs very well simulate both even (Fig. 1) and odd (Fig. 2) SCSs with the size as large as up to $\beta = 2$, within a quite wide range of the displacement amplitudes from $\alpha = -2$ up to $\alpha = 2$. Moreover, the range of values of the displacement amplitude $\alpha$, within which high fidelities $F_n^{(S\pm)}$ result, is getting wider and wider for increasing $n$. The oscillatory structure of the fidelities in the plots is caused by the $cos/sin$ dependence of the coefficients $a_k^{(\pm)}$ in Eqs. (9,10). The coefficients of the even/odd optical SCSs are shifted relative to each other by $\pi/2$ (cosine function transforms to sine with change of the phase $\varphi \to \varphi + \pi/2$). This means that when the fidelity $F_n^{(S+)}$ attains a local maximal value, the fidelity $F_n^{(S-)}$ takes a local minimum one (i.e., there is a $\pi/2$ phase-shift) under coincidental values of the parameters $\alpha$, $\beta$, and vice versa, regardless of $n$. We also numerically found the maximum values of the fidelities $F_{n,max}^{(S+)}$ (top-left) and $F_{n,max}^{(S-)}$ (top-right) as a function of $\beta$ for different values of $n$ in Fig. 3. Maximum values of the fidelities $F_{n,max}^{(S+)}$ and $F_{n,max}^{(S-)}$ follow from Figs. 1 and 2 and are determined when the displacement amplitude $\alpha$ changes with a fixed value $\beta$ of the cat's size. It is interesting to note that the maximum values of the fidelity $F_{n,max}^{(S+)}$ are observed when $n = 2, 4, 6, 8$ (i.e., $n$ is even) in the case of $\alpha = 0$; that is, when the SCQ is defined in Hilbert space with base number states (0-representation), while the maximum values of the fidelity are observed for odd values $n = 3, 5, 7, 9$ in the case of $\alpha \neq 0$ (bottom-left subfigure in Fig.3). Contrary behaviors are found for the fidelities $F_{n,max}^{(S-)}$. The maximum value of $F_{n,max}^{(S-)}$ is observed for $\alpha = 0$ in the case of $n = 3, 5, 7, 9$ but for $\alpha \neq 0$ in the case of $n = 2, 4, 6, 8$ (bottom-right subfigure in Fig.3). Summarizing the data from Figs 1-3, we list the numerical values of the size $\beta$ of the SCS and the corresponding displacement amplitude $i\alpha$ in Table 1 for which both the fidelities $F_n^{(S+)}$ and $F_n^{(S-)}$ take values greater than 0.99 $\left(F_n^{(S+)} > 0.99, F_n^{(S-)} > 0.99\right)$ for each value of $n$. A further increase in the size $\beta$ leads to the fact that the fidelities take values smaller than 0.99 $\left(F_n^{(S+)} < 0.99, F_n^{(S-)} < 0.99\right)$ for any value of $\alpha$.

|   | $F_n^{(S+)}(\beta) > 0.99$ | | $F_n^{(S-)}(\beta) > 0.99$ | |
|---|---|---|---|---|
| $n$ | $\alpha$ | $\beta$ | $\alpha$ | $\beta$ |
| 2 | 0 | 0.8615 | ±0.3409 | 0.7209 |
| 3 | ±0.328 | 1.0304 | 0 | 1.044 |
| 4 | 0 | 1.2724 | ±0.301 | 1.2267 |
| 5 | ±0.2824 | 1.4361 | 0 | 1.4574 |
| 6 | 0 | 1.6405 | ±0.266 | 1.6184 |
| 7 | ±0.2523 | 1.7933 | 0 | 1.8098 |
| 8 | 0 | 1.9715 | ±0.2404 | 1.9571 |
| 9 | ±0.2301 | 2.1131 | 0 | 2.1252 |

**Table 1.** Maximum values of $\beta$ which guarantee the fidelities exceeding 0.99 with the appropriate values of the displacement amplitude $i\alpha$. An increase in the size $\beta$ decreases the fidelities below 0.99 $\left(F_n^{(S+)} < 0.99, F_n^{(S-)} < 0.99\right)$ for any value of the displacement amplitude $\alpha$. The displacement amplitudes $\pm\alpha$ are used due to symmetry in Figs 1 and 2.

As mentioned above, the original SCSs $|\beta_+\rangle$ and $|\beta_-\rangle$ are exactly orthogonal to each other. Then, it is interesting to see to what extent the SCQs $|\Psi_n^{(S+)}\rangle$ and $|\Psi_n^{(S-)}\rangle$ are orthogonal to each other. To measure their orthogonality we plot in Fig. 4 their scalar product

$$SP_n = \langle \Psi_n^{(S-)} | \Psi_n^{(S+)} \rangle = N_n^{(S+)} N_n^{(S-)} \sum_{k=0}^n a_k^{(-)*} a_k^{(+)}, \quad (16)$$

in dependency on the parameters $\alpha$ and $\beta$. We can see from this graph that the SCQs under study become more and more orthogonal to each other as the number $n$ of terms in the superposition is increasing. The magnitude of $SP_n$ is almost completely zero in the entire range of the parameters $\alpha$ and $\beta$ for $n = 9$ which suggests that the even and odd SCQs, $|\Psi_n^{(S+)}\rangle$ and $|\Psi_n^{(S-)}\rangle$ can be regarded as orthogonal ones with $n \geq 9$, for which the fidelities are also close enough to 1, confirming the self-consistency of the approximation.

**Generation of SCQ from a two-mode entangled state.** This section proposes an optical scheme to generate the SCQs of the forms (11,12) that approximate the desired SCSs (3, 4) with high fidelity. Our scheme shown in Fig. 5 exploits the following two-mode entangled state

$$\left|\phi_n^{(\pm)}\right\rangle_{12} = \sum_{m=0}^n d_m^{(\pm)} |m\rangle_1 |n-m\rangle_2, \quad (17)$$

with the coefficients $d_m^{(\pm)}$ satisfying the normalization conditions $\sum_{m=0}^n \left|d_m^{(\pm)}\right|^2 = 1$, as the initial state. Note that in the state (17) the photon number of either mode may be any between

0 and $n$ but the total photon number of two modes is fixed to $n$. Given the coefficients $d_m^{(\pm)}$ the state (17) can be pre-produced offline in a conditional optical setup with two spontaneous parametric down converters (SPDCs) connected with each other by a set of properly-arranged beam splitters[50] (see more later). In Fig. 5 mode 1 is the main made, where the SCQ is to be born, while mode 2 is the auxiliary one, of which photons are detected ($|k\rangle\langle k|$ implies that $k$ photons are registered by a detector).

Starting from the state $\left|\phi_n^{(\pm)}\right\rangle_{12}$ two displacement operators $D_1(i\alpha)$ and $D_2(\alpha')$ (in general $\alpha \neq \alpha'$) act respectively on mode 1 and mode 2 of $\left|\phi_n^{(\pm)}\right\rangle_{12}$ resulting in

$$D_1(i\alpha)D_2(\alpha')\left|\phi_n^{(\pm)}\right\rangle_{12} = \sum_{m=0}^{n} d_m^{(\pm)} |m, i\alpha\rangle_1 |n-m, \alpha'\rangle_2, \tag{18}$$

where $|m, i\alpha\rangle = D_1(i\alpha)|m\rangle_1$ and $|n-m, \alpha'\rangle = D_2(\alpha')|n-m\rangle_2$ are the displaced number states (see (47) in Section Methods). It is worth noting that the displacement operation can be realized by mixing the target state with a strong coherent state on a highly transmissive beam splitter (HTBS)[53,54]. Then, measurement on the auxiliary mode 2 in Fig. 5 is carried out in the number states basis $\{|k\rangle; k = 0,1,2,...\}$. Using the decomposition of the displaced number state over number states as in (51), the state (18) can be rewritten as

$$D_1(i\alpha)D_2(\alpha')\left|\phi_n^{(\pm)}\right\rangle_{12} = F(\alpha') \sum_{k=0}^{\infty} N_{nk}^{(\pm)-1} \left|\Psi_{nk}^{(\pm)}\right\rangle_1 |k\rangle_2, \tag{19}$$

where the state of mode 1

$$\left|\Psi_{nk}^{(\pm)}\right\rangle_1 = N_{nk}^{(\pm)} \sum_{m=0}^{n} d_m^{(\pm)} c_{n-m,k}(\alpha') |m, i\alpha\rangle_1 \tag{20}$$

is normalized with the normalization factor

$$N_{nk}^{(\pm)} = \left(\sum_{m=0}^{n} \left|d_m^{(\pm)}\right|^2 |c_{n-m,k}(\alpha')|^2\right)^{-1/2}. \tag{21}$$

As seen from Eq. (19), conditioned on the outcome $k$ of the measurement on mode 2 (i.e., mode 2 is found in state $|k\rangle_2$ or, the same, $k$ photons of mode 2 are detected), mode 1 is immediately projected onto the state $\left|\Psi_{nk}^{(\pm)}\right\rangle_1$ of Eq. (20). Note that in (20) the subscripts $'nk'$ imply generation of a qudit of dimension $n+1$ in mode 1 which is heralded by detection of $k$ photons in mode 2, while superscripts $'\pm'$ refer to even/odd SCQs. The exponential multiplier $F(\alpha')$ in (19) is introduced in Section Method. The success probability to generate the state (20) is determined by

$$P_{nk}^{(\pm)} = F^2(\alpha') N_{nk}^{(\pm)-2} = \exp(-|\alpha'|^2) \left(\sum_{m=0}^{n} \left|d_m^{(\pm)}\right|^2 |c_{n-m,k}(\alpha')|^2\right). \tag{22}$$

Using the completeness of the displaced number states, it is straightforward to check that all the success probabilities sum to one, i.e.,

$$\sum_{k=0}^{\infty} P_{nk}^{(\pm)} = 1, \tag{23}$$

for any value of $\alpha'$ and $n$, as it should be.

Furthermore, if we impose conditions on the coefficients $d_m^{(\pm)}$ of the initial state (17) as

$$d_m^{(+)} = \frac{a_m^{(+)}/2}{c_{n-m,k}(\alpha')} N_{nk}^{(S+)'} = \frac{(i\alpha)^m \cos(\alpha\beta + m(\varphi+\pi/2))}{c_{n-m,k}(\alpha')\sqrt{m!}} N_{nk}^{(S+)'}, \tag{24}$$

or

$$d_m^{(-)} = \frac{a_m^{(-)}/2}{c_{n-m,k}(\alpha')} N_{nk}^{(S-)'} = \frac{(i\alpha)^m \sin(\alpha\beta + m(\varphi+\pi/2))}{c_{n-m,k}(\alpha')\sqrt{m!}} N_{nk}^{(S-)'}, \tag{25}$$

with the normalization factors

$$N_{nk}^{(S\pm)'} = \left(\sqrt{\sum_{m=0}^{n} \left|a_m^{(\pm)}\right|^2 / 4 |c_{n-m,k}(\alpha')|^2}\right)^{-1}, \tag{26}$$

we shall obtain the desired SCQs (11,12) whose fidelities are plotted in Figs. 1 and 2, respectively. The expressions (26) for the factors $N_{nk}^{(S\pm)'}$, which are present in the coefficients $d_m^{(\pm)}$ in Eqs. (24,25), ensure the normalization of the generated SCQs. Here we wish to note a fact that the interaction of the modes of the initial state (17) with the coherent state on the HTBS leaves its imprint in the form of the coefficients $c_{n-m,k}(\alpha')$ in the generated SCQs (i.e., the state $|\Psi_{nk}^{(\pm)}\rangle$ in (20)). It can serve inherent irreducible feature of the DV-CV interaction. The success probabilities to generate the SCQs (11,12) are given by

$$P_{nk}^{(S\pm)} = F^2 N_{nk}^{(S\pm)'2} / N_n^{(S\pm)2}, \tag{27}$$

which depend on three parameters $\alpha'$, $\alpha$ and $\beta$. We built the maximum values of the success probabilities that can be obtained for certain values of the auxiliary parameter $\alpha'$ in dependence on $\alpha$ and $\beta$. Maximal success probabilities $P_{n0}^{(S+)}$ and $P_{n1}^{(S+)}$ are shown in Figs 6 and 7, while quantities $P_{n0}^{(S-)}$ and $P_{n1}^{(S-)}$ are displayed in Figs 8 and 9, respectively. These values also depend on the number of terms in generated superposition $n$ and on the number of the registered photons $k$. The general tendency is that the approximation under consideration here is better for a larger $n$ but the corresponding maximal success probability decreases with increasing $n$.

At this point, we briefly address on a possibility to generate the two-mode entangled state $|\phi_n^{(\pm)}\rangle_{12}$ in (17), following the work[50]. For concreteness, let us consider the state $|\phi_n^{(+)}\rangle_{12}$ which can be reformulated in terms of the bosonic modal creation operators $a_1^+$ and $a_2^+$ as

$$|\phi_n^{(+)}\rangle_{12} = \sum_{m=0}^{n} \frac{d_m^{(\pm)}}{\sqrt{m!(n-m)!}} a_1^{+m} a_2^{+(n-m)} |0\rangle_1 |0\rangle_2, \tag{28}$$

with $|0\rangle_1|0\rangle_2$ the two-mode vacuum state. If we pull $a_2^{+n}$ out of the sum and introduce a formal variable $z = a_1^+/a_2^+$, then (28) reads

$$|\phi_n^{(+)}\rangle_{12} = a_2^{+n} f(z) |0\rangle_1 |0\rangle_2, \tag{29}$$

where

$$f(z) = \sum_{m=0}^{n} \frac{d_m^{(\pm)}}{\sqrt{m!(n-m)!}} z^m \tag{30}$$

is a nonconstant single-variable $n^{th}$ order polynomial in $z$ with complex coefficients. According to the fundamental theorem of algebra, the polynomial $f(z)$ above can always be factorized out as

$$f(z) = \frac{d_n^{(+)}}{\sqrt{n!}} \prod_{m=0}^{n} \left(z - z_m^{(+)}\right), \tag{31}$$

with $z_m^{(+)}$ solutions of the equation $f(z) = 0$. Putting (31) back into (29) yields

$$|\phi_n^{(+)}\rangle_{12} = \frac{d_n^{(+)}}{\sqrt{n!}} \left[\prod_{m=0}^{n} \left(a_1^+ - z_m^{(+)} a_2^+\right)\right] |0\rangle_1 |0\rangle_2. \tag{32}$$

By changing the variables $z_m^{(+)} \to -r_m^{(+)}/t_m^{(+)}$, with $r_m^{(+)}$ and $t_m^{(+)}$ such that $|r_m^{(+)}|^2 + |t_m^{(+)}|^2 = 1$, we get

$$|\phi_n^{(+)}\rangle_{12} = \frac{d_n^{(+)}}{\sqrt{n!} \prod_{m=0}^{n} t_m^{(+)}} \left[\prod_{m=0}^{n} \left(t_m^{(+)} a_1^+ + r_m^{(+)} a_2^+\right)\right] |0\rangle_1 |0\rangle_2. \tag{33}$$

The parameters $t_m^{(+)}$ and $r_m^{(+)}$ can be treated as transmission and reflection coefficients of a beam splitter which are determined by $z_m^{(+)}$ in the following manner

$$t_m^{(+)} = \frac{1}{\sqrt{1 + |z_m^{(+)}|^2}}, \tag{34}$$

$$r_m^{(+)} = -\frac{z_m^{(+)}}{\sqrt{1+\left|z_m^{(+)}\right|^2}}.\tag{35}$$

Because (33) is a product of terms that are linear in the modal creation operators acting on the two-mode vacuum state, such states can be generated by a heralded scheme proposed in Ref.[50]. The scheme starts from two two-mode squeezed states produced by two independent SPDCs. Each squeezed state has a signal mode and an idler mode. First, each idler mode is splitted into $n$ modes with an equal weight by a set of $n-1$ unbalanced beam splitters with proper transmission and reflection coefficients. Then, the splitted modes from one idler mode are correspondingly superposed with those from the other idler mode on $n$ beam splitters with transmission and reflection coefficients $(t_1^{(+)}, r_1^{(+)})$, $(t_1^{(+)}, r_1^{(+)})$, ... and $(t_n^{(+)}, r_n^{(+)})$, respectively. Behind each such beam splitter there is a photo-detector. If each detector registers a photon, then the two signal modes are projected onto the state $\left|\phi_n^{(+)}\right\rangle_{12}$. The same procedures apply to generation of the state $\left|\phi_n^{(-)}\right\rangle_{12}$. The state generation process described above is probabilistic but this does not matter since $\left|\phi_n^{(\pm)}\right\rangle_{12}$ are to be generated offline and only after they are successfully generated we shall turn to the problem of generation of our SCQs as in Fig. 5. Because analytically finding the solutions $\{z_m^{(\pm)}; m = 0,1,...,n\}$ for specific coefficients $\{d_m^{(\pm)}; m = 0,1,2,...,n\}$ is generally difficult, we, for illustration, carry out numerical calculation for the cases of $n = 3, 6, 9$ and $12$ and some given values of the displacement amplitudes $\alpha$ and $\alpha'$ (see Fig. 5) for which the fidelities $F_n^{(S\pm)} > 0.99$. The calculated values of $t_m^{(\pm)}$ and $r_m^{(\pm)}$ are collected in Tables 2 and 3. As can be seen from the Tables, high-fidelity SCQs of large size ($\beta \geq 2$) can be produced in the case of relatively large $n$ (say, $n \geq 9$). Since the concerned optical devices (SPDCs, beam splitters, phase shifters, photo-detectors) are available within the current technologies and the necessary numerical calculation is not formidable with the help of modern computing facilities, the presented production of large-size optical Schrödinger cat states seems quite efficient.

| | $\left|\Psi_3^{(S+)}\right\rangle$ | $\left|\Psi_6^{(S+)}\right\rangle$ | $\left|\Psi_9^{(S+)}\right\rangle$ | $\left|\Psi_{12}^{(S+)}\right\rangle$ |
|---|---|---|---|---|
| $\beta$ | 1.03 | 1.64 | 2.12 | 2.55 |
| $\alpha$ | 0.328 | 0 | 0.23 | 0 |
| $\alpha'$ | 1.426 | 1.805 | 2.048 | 2.248 |
| $P_{n0}^{(S+)}$ | 0.20 | 0.12 | 0.08 | 0.06 |
| $t_1^{(+)}$ | 0.755 | 0.582 | 0.574 | 0.563 |
| $r_1^{(+)}$ | $-i0.656$ | $0.814 exp(i0.399\pi)$ | $0.818 exp(-i0.299\pi)$ | $0.826 exp(-i0.286\pi)$ |
| $t_2^{(+)}$ | 0.634 | 0.582 | 0.577 | 0.563 |
| $r_2^{(+)}$ | $i0.773$ | $0.814 exp(-i0.399\pi)$ | $0.817 exp(i0.363\pi)$ | $0.826 exp(i0.286\pi)$ |
| $t_3^{(+)}$ | 0.547 | 0.883 | 0.698 | 0.663 |
| $r_3^{(+)}$ | $-i0.837$ | $i0.469$ | $0.716 exp(-i0.469\pi)$ | $0.749 exp(-i0.428\pi)$ |
| $t_4^{(+)}$ | | 0.883 | 0.97 | 0.663 |
| $r_4^{(+)}$ | | $-i0.469$ | $-i0.242$ | $0.749 exp(i0.428\pi)$ |
| $t_5^{(+)}$ | | 0.582 | 0.904 | 0.964 |
| $r_5^{(+)}$ | | $0.814 exp(-i0.601\pi)$ | $i0.428$ | $i0.264$ |
| $t_6^{(+)}$ | | 0.582 | 0.674 | 0.964 |
| $r_6^{(+)}$ | | $0.814 exp(i0.601\pi)$ | $i0.739$ | $-i0.264$ |
| $t_7^{(+)}$ | | | $0.698$ | $0.774$ |
| $r_7^{(+)}$ | | | $0.716 exp(-i0.531\pi)$ | $-i0.633$ |
| $t_8^{(+)}$ | | | $0.577$ | $0.774$ |
| $r_8^{(+)}$ | | | $0.817 exp(i0.637\pi)$ | $i0.633$ |
| $t_9^{(+)}$ | | | $0.574$ | $0.663$ |
| $r_9^{(+)}$ | | | $0.818 exp(-i0.701\pi)$ | $0.749 exp(i0.572\pi)$ |
| $t_{10}^{(+)}$ | | | | $0.663$ |
| $r_{10}^{(+)}$ | | | | $0.749 exp(-i0.572\pi)$ |
| $t_{11}^{(+)}$ | | | | $0.563$ |
| $r_{11}^{(+)}$ | | | | $0.826 exp(i0.714\pi)$ |
| $t_{12}^{(+)}$ | | | | $0.563$ |
| $r_{12}^{(+)}$ | | | | $0.826 exp(-i0.714\pi)$ |

**Table 2.** Values of the beam splitter parameters $t_i^{(+)}$, $r_i^{(+)}$ use of which in optical scheme[50] ensures the generation of needed two-mode entangled state (17). Application of the displacement operators with amplitudes $\alpha$ and $\alpha'$ in Fig. 5 enables us to generate SCQ that most closely match the properties of the even SCS of the corresponding size $\beta$ with fidelity $F_n^{(S+)} > 0.99$.

| | $\left|\Psi_3^{(S-)}\right\rangle$ | $\left|\Psi_6^{(S-)}\right\rangle$ | $\left|\Psi_9^{(S-)}\right\rangle$ | $\left|\Psi_{12}^{(S-)}\right\rangle$ |
|---|---|---|---|---|
| $\beta$ | 1.04 | 1.62 | 2.13 | 2.54 |
| $\alpha$ | 0 | 0.266 | 0 | 0.205 |
| $\alpha'$ | 1.265 | 1.77 | 2.042 | 2.248 |
| $P_{n0}^{(S-)}$ | 0.25 | 0.13 | 0.08 | 0.06 |
| $t_1^{(-)}$ | 1 | 0.575 | 0.574 | 0.562 |
| $r_1^{(-)}$ | 0 | $0.818 exp(-i0.356\pi)$ | $0.819 exp(i0.331\pi)$ | $0.827 exp(-i0.262\pi)$ |
| $t_2^{(-)}$ | 0.473 | 0.596 | 0.574 | 0.563 |
| $r_2^{(-)}$ | $i0.881$ | $0.802 exp(i0.449\pi)$ | $0.819 exp(-i0.331\pi)$ | $0.826 exp(i0.311\pi)$ |
| $t_3^{(-)}$ | 0.473 | 0.989 | 1 | 0.663 |
| $r_3^{(-)}$ | $-i0.881$ | $i0.149$ | 0 | $0.748 exp(-i0.403\pi)$ |
| $t_4^{(-)}$ | | 0.737 | 0.81 | 0.665 |
| $r_4^{(-)}$ | | $-i0.676$ | $i0.586$ | $0.747 exp(i0.448\pi)$ |
| $t_5^{(-)}$ | | 0.596 | 0.81 | 0.996 |
| $r_5^{(-)}$ | | $0.802 exp(i0.551\pi)$ | $-i0.586$ | $i0.091$ |
| $t_6^{(-)}$ | | 0.575 | 0.659 | 0.909 |
| $r_6^{(-)}$ | | $0.818 exp(-i0.644\pi)$ | $i0.752$ | $-i0.417$ |
| $t_7^{(-)}$ | | | 0.659 | 0.842 |
| $r_7^{(-)}$ | | | $-i0.752$ | $i0.540$ |
| $t_8^{(-)}$ | | | 0.574 | 0.724 |
| $r_8^{(-)}$ | | | $0.819 exp(i0.669\pi)$ | $-i0.690$ |
| $t_9^{(-)}$ | | | 0.574 | 0.665 |
| $r_9^{(-)}$ | | | | $0.819 exp(-i0.669\pi)$ | $0.747 exp(i0.552\pi)$ |
| $t_{10}^{(-)}$ | | | | | $0.663$ |
| $r_{10}^{(-)}$ | | | | | $0.748 exp(-i0.597\pi)$ |
| $t_{11}^{(-)}$ | | | | | $0.563$ |
| $r_{11}^{(-)}$ | | | | | $0.826 exp(i0.689\pi)$ |
| $t_{12}^{(-)}$ | | | | | $0.562$ |
| $r_{12}^{(-)}$ | | | | | $0.827 exp(-i0.738\pi)$ |

**Table 3.** Values of the beam splitter parameters $t_i^{(-)}$, $r_i^{(-)}$ use of which in optical scheme[50] ensures the generation of needed two-mode entangled state (17). Application of the displacement with amplitudes $\alpha$ and $\alpha'$ in Fig. 5 enables us to generate SCQ that most closely match the properties of the odd SCS of the corresponding size $\beta$ with fidelity $F_n^{(S-)} > 0.99$.

**Generation of SCQ from Fock states.** In this section we propose another scheme to generate the SCQs, $|\Psi_n^{(S\pm)}\rangle$ in Eqs. (11,12), using $m+1 \geq 2$ photon Fock states $|k_0\rangle_0$, $|k_1\rangle_1, \ldots, |k_m\rangle_m$ with $k_0, k_1, \ldots, k_m \geq 0$, as the inputs. Our scheme is sketched in Fig. 10 which consists of $m$ beam splitters, $m+1$ displacement operations and $m$ photo-detectors. For any given $m \geq 1$, if neither of the $m$ detectors clicks, the output state of the form (See Section Methods)

$$|\Omega_n^{(m)}\rangle_0 = N_n^{(m)} D_0(i\alpha) \prod_{k=1}^n D_0(\beta_k^{(m)*}) a^+ D_0^\dagger(\beta_k^{(m)*}) |0\rangle_0 \quad (36)$$

is generated, where

$$n = k_0 + k_1 + \cdots + k_m \quad (37)$$

and $\{\beta_k^{(m)}; k = 1, 2, \ldots, n\}$ being functions of the parameters $t_1, r_1, t_2, r_2, \ldots, t_m, r_m$ of the beam splitters and $\alpha_1, \alpha_2, \ldots, \alpha_m$ of the displacement operators. The state $|\Omega_n^{(m)}\rangle$ can be made coincident with the desired SCQs $|\Psi_n^{(S\pm)}\rangle$ by properly choosing the involved parameters. Namely, since the SCQs of Eqs. (11,12) can be rewritten as

$$|\Psi_n^{(S\pm)}\rangle = D(i\alpha) \sum_{k=0}^n c_k^{(S\pm)} |k\rangle, \quad (38)$$

with $c_k^{(S\pm)} = N_n^{(S\pm)} a_k^{(\pm)}/2$, it can also be expressed in the form (36)[39], i.e.,

$$|\Psi_n^{(S\pm)}\rangle = D(i\alpha) \frac{c_n^{(S\pm)}}{\sqrt{n!}} \prod_{k=1}^n D(\gamma_k^{(\pm)*}) a^+ D^+(\gamma_k^{(\pm)*}) |0\rangle, \quad (39)$$

where $\{\gamma_k^{(\pm)}; k = 1, 2, \ldots, n\}$ are the $n$ roots of the polynomial

$$\sum_{k=0}^n \frac{c_k^{(S\pm)} \sqrt{n!}}{c_n^{(S\pm)} \sqrt{k!}} \gamma^{(\pm)k} = 0. \quad (40)$$

Note that each root $\gamma_k^{(\pm)}$ depends on the coefficients $\{c_j^{(S\pm)}; j = 0,1, \ldots, n\}$ of the desired SCQ (38). It follows from comparing (36) and (39) that, for a given set of $\{c_j^{(S\pm)}\}$, the scheme's parameters $t_1, r_1, \alpha_1, t_2, r_2, \alpha_2, \ldots, t_m, r_m, \alpha_m$ can be chosen such that to satisfy the equations

$$N_n^{(m)} = \frac{c_n^{(S\pm)}}{\sqrt{n!}} \quad (41)$$

and

$$\beta_k^{(m)} = \gamma_k^{(\pm)} \quad \forall k. \quad (42)$$

If so, $|\Omega_n^{(m)}\rangle$ becomes $|\Psi_n^{(S\pm)}\rangle$, implying generation of the desired SCQ from Fock states $|k_0\rangle_0, |k_1\rangle_1, \ldots, |k_m\rangle_m$ by our scheme in Fig. 10.

For illustration, for the $m = 1$ case our calculations yield (see Section Methods)

$$N_n^{(1)} = \frac{1}{\sqrt{P_n^{(1)}}} \frac{(t_1)^{k_0}(-r_1^*)^{k_1}}{\sqrt{k_0!k_1!}} exp\left(-\frac{|\alpha_1|^2}{2}\right), \tag{43}$$

where $n = k_0 + k_1$, $P_n^{(1)}$ the success probability (see Section Methods) and

$$\beta_k^{(1)} = \begin{cases} \frac{r_1}{t_1}\alpha_1^*; & k \in [1, k_0] \\ -\frac{t_1^*}{r_1^*}\alpha_1^*; & k \in [k_0 + 1, k_0 + k_1] \end{cases}. \tag{44}$$

As for the $m = 2$ case we arrive at (see Section Methods)

$$N_n^{(2)} = \frac{1}{\sqrt{P_n^{(2)}}} \frac{(t_1t_2)^{k_0}(-r_1^*t_2)^{k_1}(-r_2^*)^{k_2}}{\sqrt{k_0!k_1!k_2!}} exp\left(-\frac{|\alpha_1|^2+|\alpha_2|^2}{2}\right), \tag{45}$$

where $n = k_0 + k_1 + k_2$, $P_n^{(2)}$ the success probability (see Section Methods), and

$$\beta_k^{(2)} = \begin{cases} \frac{t_1r_2\alpha_2^*+r_1\alpha_1^*}{t_1t_2}; & k \in [1, k_0] \\ \frac{r_1^*r_2\alpha_2^*-t_1^*\alpha_1^*}{r_1^*t_2}; & k \in [k_0 + 1, k_0 + k_1] \\ \frac{t_2^*\alpha_2^*}{-r_2^*}; & k \in [k_0 + k_1 + 1, k_0 + k_1 + k_2] \end{cases}. \tag{46}$$

From the above description, we see that our scheme works for any $m \geq 1$. It seems that the smaller value of $m$ (i.e., the lesser the number of used beam splitters/displacement operators/detectors) the better the scheme with respect to the devices consumption. However, for a given $n$, a smaller value of $m$ is accompanied by larger values of $k_0, k_1, \ldots, k_m$ to meet the requirement (37). Also, the described scheme is probabilistic because of its post-selection procedure. In fact, there may be a wide range of choice of possible parameters that satisfy the equations (41) and (42) with high accuracy; yet each choice leads to a different success probability.

In what follows, for concreteness, let us deal with generation of the SCQs $|\Psi_{10}^{(S\pm)}\rangle$ of size $\beta = 2$ for three sets of $(m, k_0, k_1, \ldots, k_m)$:

(i)          $m = 3, k_0 = 4, k_1 = k_2 = k_3 = 2$,
(ii)        $m = 4, k_0 = 2, k_1 = k_2 = k_3 = k_4 = 2$,
(iii)      $m = 5, k_0 = 0, k_1 = k_2 = k_3 = k_4 = k_5 = 2$.

The results of numerical calculations are listed in Tables 4, 5 and 6, respectively.

|  | $|\Psi_{10}^{(S+)}\rangle$ | $|\Psi_{10}^{(S-)}\rangle$ |
|---|---|---|
| $F_{10}^{(3)}$ | 0.98 | 0.961 |
| $\alpha$ | $-0.35$ | 0.44 |
| $P_{10}^{(3)}$ | 0.0015 | 0.0071 |
| $\alpha_1$ | $1.657 \cdot exp(i0.485\pi)$ | $1.999 \cdot exp(i0.161\pi)$ |
| $\alpha_2$ | $0.274 \cdot exp(i0.475\pi)$ | $-0.270$ |
| $\alpha_3$ | $1.176 \cdot exp(i0.876\pi)$ | $1.164 \cdot exp(i0.784\pi)$ |
| $t_1$ | $0.614 \cdot exp(i0.010\pi)$ | $0.732 \cdot exp(i0.161\pi)$ |
| $t_2$ | $0.684 \cdot exp(i0.600\pi)$ | $0.760 \cdot exp(i1.216\pi)$ |
| $t_3$ | $0.664 \cdot exp(i1.376\pi)$ | $0.690 \cdot exp(i1.284\pi)$ |

**Table 4.** Numerical results of the chosen parameters for generation of the SCQs $|\Psi_{10}^{(S\pm)}\rangle$ with size $\beta = 2$ for the case (i), i.e., when $m = 3$, $k_0 = 4, k_1 = k_2 = k_3 = 2$. $F_{10}^{(3)}$ and $P_{10}^{(3)}$ are the corresponding fidelity and success probability.

| | $|\Psi_{10}^{(S+)}\rangle$ | $|\Psi_{10}^{(S-)}\rangle$ |
|---|---|---|
| $F_{10}^{(4)}$ | 0.985 | 0.972 |
| $\alpha$ | $-0.47$ | $-0.14$ |
| $P_{10}^{(4)}$ | 0.0007 | 0.0017 |
| $\alpha_1$ | $1.214 \cdot exp(-i0.46\pi)$ | $1.405 \cdot exp(i0.258\pi)$ |
| $\alpha_2$ | $0.841 \cdot exp(i0.175\pi)$ | $1.026 \cdot exp(-i0.905\pi)$ |
| $\alpha_3$ | 0 | $0.091 \cdot exp(i0.223\pi)$ |
| $\alpha_4$ | $1.222 \cdot exp(i0.946\pi)$ | $1.204 \cdot exp(i0.543\pi)$ |
| $t_1$ | $0.755 \cdot exp(i0.683\pi)$ | $0.603 \cdot exp(i0.371\pi)$ |
| $t_2$ | $0.798 \cdot exp(i0.056\pi)$ | $0.847 \cdot exp(i0.158\pi)$ |
| $t_3$ | $0.531 \cdot exp(i1.935\pi)$ | $0.595 \cdot exp(i1.314\pi)$ |
| $t_4$ | $0.829 \cdot exp(i0.446\pi)$ | 0.917 |

**Table 5.** Numerical results of the chosen parameters for generation of the SCQs $|\Psi_{10}^{(S\pm)}\rangle$ with size $\beta = 2$ for the case (ii), i.e., when $m = 4$, $k_0 = 2$, $k_1 = k_2 = k_3 = k_4 = 2$. $F_{10}^{(4)}$ and $P_{10}^{(4)}$ are the corresponding fidelity and success probability.

| | $|\Psi_{10}^{(S+)}\rangle$ | $|\Psi_{10}^{(S-)}\rangle$ |
|---|---|---|
| $F_{10}^{(5)}$ | 0.975 | 0.973 |
| $\alpha$ | $-0.2$ | $-0.28$ |
| $P_{10}^{(5)}$ | 0.0008 | 0.0012 |
| $\alpha_1$ | 0 | $0.034 \cdot exp(i0.940\pi)$ |
| $\alpha_2$ | $1.216 \cdot exp(i0.588\pi)$ | $1.585 \cdot exp(-i0.686\pi)$ |
| $\alpha_3$ | $0.738 \cdot exp(-i0.982\pi)$ | $0.948 \cdot exp(i0.310\pi)$ |
| $\alpha_4$ | $0.99 \cdot exp(i0.390\pi)$ | $1.401 \cdot exp(-i0.021\pi)$ |
| $\alpha_5$ | $1.414 \cdot exp(-i0.277\pi)$ | $0.295 \cdot exp(i0.037\pi)$ |
| $t_1$ | $0.468 \cdot exp(i1.116\pi)$ | $0.444 \cdot exp(i1.321\pi)$ |
| $t_2$ | $0.414 \cdot exp(i0.570\pi)$ | $0.611 \cdot exp(i0.037\pi)$ |
| $t_3$ | $0.506 \cdot exp(i0.628\pi)$ | $0.702 \cdot exp(i1.084\pi)$ |
| $t_4$ | $0.868 \cdot exp(i1.668\pi)$ | $0.876 \cdot exp(i0.087\pi)$ |
| $t_5$ | $0.754 \cdot exp(i1.223\pi)$ | $0.767 \cdot exp(i0.537\pi)$ |

**Table 6.** Numerical results of the chosen parameters for generation of the SCQs $|\Psi_{10}^{(S\pm)}\rangle$ with size $\beta = 2$ for the case (iii), i.e., when $m = 5$, $k_0 = 0, k_1 = k_2 = k_3 = k_4 = k_5 = 2$. $F_{10}^{(5)}$ and $P_{10}^{(5)}$ are the corresponding fidelity and success probability.

As can be seen from Tables 4, 5 and 6, generation of SCQs with size as large as $\beta = 2$ is possible in all the three cases with high enough fidelity whose values range from 0.961 up to 0.985. The obtained success probabilities to generate the SCQs are quite small but these are typical for this kind of state generation. Generally speaking, none of the proposed interpretations provides significant advantages over each other. Nevertheless, the data from the Tables reveal the correctness of the proposed scheme which allows to realize SCQs of large size starting from an original tensor product of Fock states. Since the SCQ generation in the case (i) requires a smaller number of beam splitters, displacement operators and photo-detectors than in the cases (ii) and (iii), this case can be considered as more effective from an experimental point of view. Finally, to confirm the correctness of the proposed scheme which is rather complicated from a numerical point of view, we use the numerical values of the amplitudes of the superposition (36) to construct Wigner functions of the generated (left subfigures) and those of the genuine SCS $|\beta_+\rangle$ with $\beta = 2$ (right subfigures). SCQs and compare them with the Wigner functions of the corresponding genuine SCSs. In Fig. 11 we use the numerical data in Table 4 to plot Wigner functions of the SCQ $|\Psi_{10}^{(S+)}\rangle$ (left subfigures) and those of the genuine SCS $|\beta_+\rangle$ with $\beta = 2$ (right subfigures). The fidelity calculated for the two Wigner functions gives the following value $F_{10} = 0.980140336082$ which completely coincides with the value of the fidelity presented in Table 4. Also, we show in Fig. 12 Wigner functions of the SCQ $|\Psi_{10}^{(S-)}\rangle$ (left subfigures) and the genuine SCS $|\beta_-\rangle$ with $\beta = 2$ (right subfigures). Again, the fidelity calculated using the Wigner functions gives the value $F_{10} = 0.961285449744$ which is the same as that presented in Table 4. Note that the Wigner functions $W_{SCQ}$ and $W_{SCS}$ of both the generated SCQ and the genuine SCS exhibit areas of negativity (i.e., areas for which $W_{SCQ}, W_{SCS} < 0$), which is a specific feature to ensure nonclassicality of the states of concern. This observation and the full coincidence of the values of fidelities calculated by two different ways allow us to positively judge the relevance of the proposed scheme to generate large-size SCQs from Fock states.

## Discussion

We have considered novel ways to generate displaced qudits, called Schrodinger cat qudits, which may approximate Schrodinger cat states of big size with high fidelity. First, we developed a theory of $\alpha$-representation of the Schrodinger cat states (Eqs. (9, 10)), where the quantity $\alpha$ takes pure imaginary values. The amplitudes of even and odd Schrodinger cat states are shifted relative to each other by $\pi/2$. Therefore, the division of the states onto even and odd can be made only in number states base (0-representation). These states have both even and odd amplitudes in any other Hilbert space defined by the displacement amplitude $\alpha$. Schrodinger cat qudits are determined in an $(n + 1)$-dimensional Hilbert space with displaced base elements (47) shifted by quantity $\alpha$ on phase plane regarding the number states. Schrodinger cat qudits give maximal fidelity with exact Schrodinger cat states for any values of the displacement amplitude $\alpha$. The more the number of terms $n$ in the displaced qudit we take, the higher fidelity we can approximate the Schrodinger cat states of large size (see Figs. 1-3). It is interesting to note that even and odd Schrodinger cat qudits have maximum fidelity in 0-representation for $n$ being even and odd, respectively.

Then, we propose possible methods of generating Schrodinger cat qudits. One method is based on a two-mode entangled state (17) containing $n$ photons in total. The amplitudes of this state follow from Eqs. (24, 25) and depend on both Schrodinger cat states amplitudes (9, 10) and decomposition coefficients (51). The generation of even/odd Schrodinger cat qudits in optical scheme in Fig. 5 can be performed with a fairly high probability of success (Figs. (6-9)). It is shown[50] that the two-mode entangled $n$-photon state can be realized with the help of two SPDCs and a system of the beam splitters with parameters (Eqs. (34, 35)) determined by the roots of the equation (31). After such an entangled state (17) is produced offline like quantum channel in[55], either even or odd Schrodinger cat qudits can be generated using the amplitude displacement both in the main and auxiliary modes with the subsequent registration of a specific measurement outcome in number state basis. Potentially, this scheme allows one to realize Schrodinger cat qudits with a size greater than or equal to two ($\beta \geq 2$) with an increase in the number $n$ of photons used. Despite the simplicity of implementation of the conditional Schrodinger cat qudit generation, this scheme requires quantum channel[55], realization of which may require great efforts. In order to seek for more possibilities of implementing large-size Schrodinger cat qudits, we proposed another scheme without using the initial two-mode entangled $n$-photon state. Instead, $m+1$ ($m \geq 1$) photon number states are used as the input states. With the help of photo-detectors and linear optics devices with properly chosen parameters and arranged as in Fig. 10, large-size Schrodinger cat qudits with high fidelity with the desired Schrodinger cat states can be obtained if no detectors click. The relevance of the method of generation of the desired Schrodinger cat states from photon Fock states is confirmed by means of Wigner functions.

## Methods

**Derivation of $\alpha$-representation of the SCS.** The displaced number states[17] are defined through a unitary operator called the displacement operator $D(\alpha) = exp(\alpha a^+ - \alpha^* a)$ acting on a Fock state $|n\rangle$ as

$$|n, \alpha\rangle = D(\alpha)|n\rangle, \qquad (47)$$

where $\alpha$ is a complex number in general, $a$ ($a^+$) is the bosonic annihilation (creation) operator[56]. Set of the displaced number states of light

$$\{|n, \alpha\rangle, n = 0,1,2, \ldots, \infty\} \qquad (48)$$

is complete for a given $\alpha$. Therefore, any state can be decomposed in terms of the displaced number states with respective coefficients. We name such decomposition $\alpha$-representation. In particular, for $\alpha = 0$, the 0-representation is nothing else but the decomposition in terms of the number states. So, the 0-representation of the even and odd SCSs in Eqs. (1, 2) can be written as column-vectors with infinite number of elements as

$$|\beta_+\rangle = \begin{bmatrix} a_0^{(+)} \\ a_1^{(+)} \\ a_2^{(+)} \\ a_3^{(+)} \\ a_4^{(+)} \\ a_5^{(+)} \\ a_6^{(+)} \\ a_7^{(+)} \\ \vdots \end{bmatrix} = G_+ \begin{bmatrix} 1 \\ 0 \\ \beta^2/\sqrt{2!} \\ 0 \\ \beta^4/\sqrt{4!} \\ 0 \\ \beta^6/\sqrt{6!} \\ 0 \\ \vdots \end{bmatrix}, \qquad (49)$$

$$|\beta_-\rangle = \begin{bmatrix} a_0^{(-)} \\ a_1^{(-)} \\ a_2^{(-)} \\ a_3^{(-)} \\ a_4^{(-)} \\ a_5^{(-)} \\ a_6^{(-)} \\ a_7^{(-)} \\ \vdots \end{bmatrix} = G_- \begin{bmatrix} 0 \\ \beta \\ 0 \\ \beta^3/\sqrt{3!} \\ 0 \\ \beta^5/\sqrt{5!} \\ 0 \\ \beta^7/\sqrt{7!} \\ \vdots \end{bmatrix}, \quad (50)$$

where the normalization factors are $G_\pm = 2N_\pm(\beta)exp(-|\beta|^2/2)$. Note also that $|\beta_+\rangle$ contains only amplitudes proportional to the size $\beta^{2k}$, while $|\beta_-\rangle$ is realized with amplitudes proportional to the size $\beta^{2k+1}$. The 0-representation of the displaced number state itself is

$$|k,\alpha\rangle = F(\alpha) \sum_{n=0}^{\infty} c_{kn}(\alpha) |n\rangle, \quad (51)$$

where $F(\alpha) = exp(-|\alpha|^2/2)$ is the normalization factor and the matrix elements $c_{kn}(\alpha)$ are the decomposition coefficients of the displaced number state $|k,\alpha\rangle$ over the number states $|n\rangle^{52}$, which satisfy the condition $F(\alpha)^2 \sum_{m=0}^{\infty} |c_{kn}(\alpha)|^2 = 1$ because $|k,\alpha\rangle$ is normalized to 1. These coefficients are the matrix elements of the transformation matrix $U^{52}$. To get rid of the tedious calculations associated with the multiplication of the unitary infinite transformation matrix by the column vector[52], we directly obtain the amplitudes of the even/odd SCS in arbitrary $\alpha$-representation. Let us do the mathematical calculations for amplitudes of even SCS (1) in infinite Hilbert space of the displaced number states $|k,\alpha\rangle$. The amplitude $a_k^{(+)}$ of even SCS in $\alpha$-representation can be calculated as

$$a_k^{(+)} = \langle k,\alpha|even\rangle = N_+(\langle k,\alpha|-\beta\rangle + \langle k,\alpha|\beta\rangle) = \\ N_+(\langle k|D(-\alpha)D(-\beta)|0\rangle + \langle k|D(-\alpha)D(\beta)|0\rangle), \quad (52)$$

due to completeness of the base displaced number states. Using the operator theorem[56],

$$D(\alpha)D(\beta) = D(\alpha+\beta)exp\left(\frac{\alpha\beta^* - \alpha^*\beta}{2}\right) = D(\alpha+\beta)exp(iIm(\alpha\beta^*)), \quad (53)$$

applied to the displacement operators, we have from (A6)

$$a_k^{(+)} = N_+\left(\langle k|-\alpha-\beta\rangle exp\left(\frac{\alpha\beta^* - \alpha^*\beta}{2}\right) + \langle k|-\alpha+\beta\rangle exp\left(\frac{-\alpha\beta^* + \alpha^*\beta}{2}\right)\right) = \\ N_+\left(e^{-|-\alpha-\beta|^2/2}\frac{(-\alpha-\beta)^k}{\sqrt{k!}}exp\left(\frac{\alpha\beta^* - \alpha^*\beta}{2}\right) + e^{-|-\alpha+\beta|^2/2}\frac{(-\alpha+\beta)^k}{\sqrt{k!}}exp\left(\frac{-\alpha\beta^* + \alpha^*\beta}{2}\right)\right). \quad (54)$$

Finally, we need to group the phase factors

$$exp\left(\frac{1}{2}(-(-\alpha-\beta)(-\alpha^* - \beta^*) + \alpha\beta^* - \alpha^*\beta)\right) = \\ exp\left(\frac{1}{2}(-\alpha\alpha^* - \alpha\beta^* - \beta\alpha^* - \beta\beta^* + \alpha\beta^* - \alpha^*\beta)\right) = \\ exp\left(-\frac{1}{2}(|\alpha|^2 + |\beta|^2) - \alpha^*\beta\right), \quad (55)$$

in the first term of (A8) and

$$exp\left(\frac{1}{2}(-(-\alpha+\beta)(-\alpha^* + \beta^*) - \alpha\beta^* + \alpha^*\beta)\right) = \\ exp\left(\frac{1}{2}(-\alpha\alpha^* + \alpha\beta^* + \beta\alpha^* - \beta\beta^* - \alpha\beta^* + \alpha^*\beta)\right) = \\ exp\left(-\frac{1}{2}(|\alpha|^2 + |\beta|^2) + \alpha^*\beta\right), \quad (56)$$

in the second term of (A7). Inserting all the phase factors into (A8), we obtain

$$a_k^{(+)} = \frac{N_+}{\sqrt{k!}} \exp\left(-\frac{1}{2}(|\alpha|^2 + |\beta|^2)\right)\left((-\alpha - \beta)^k \exp(-\alpha^*\beta) + (-\alpha + \beta)^k \exp(\alpha^*\beta)\right). \quad (57)$$

If we put the common factor $N_+ exp\left(-\frac{|\alpha|^2+|\beta|^2}{2}\right) = N_+ exp\left(-\frac{a^2}{2}\right)$ out of the bracket, we get the superposition (3) whose coefficients $a_k^{(+)}$ are now determined by the formula (5). In similar manner, the amplitudes $a_k^{(-)}$ of odd SCS can be derived from relation

$$a_k^{(-)} = \langle k, \alpha | odd \rangle. \quad (58)$$

Now, consider derivation of the formulas (9, 100). We turn to the polar coordinates, given that $\beta > 0$ and the displacement amplitude $\alpha = i\alpha$ is pure imaginary quantity. Then, we have $(-\alpha - \beta)^k = a^k exp(ik\varphi)exp(ik\pi)$ and $(-\alpha + \beta)^k = a^k exp(-ik\varphi)$, where the angle on phase space is determined in Section 2. Substituting the expressions into formulas for $a_k^{(\pm)}$, we obtain

$$a_k^{(+)} = \frac{N_+}{\sqrt{k!}} \exp\left(-\frac{a^2}{2}\right) a^k \left(exp(i\alpha\beta + ik\varphi + ik\pi) + exp(-i\alpha\beta - ik\varphi)\right) =$$
$$\frac{N_+}{\sqrt{k!}} \exp\left(-\frac{a^2}{2}\right) a^k exp(ik\pi/2) \left(exp(i\alpha\beta + ik\varphi + ik\pi/2) + exp(-i\alpha\beta - ik\varphi - ik\pi/2)\right) =$$
$$\frac{N_+}{\sqrt{k!}} 2(ia)^k \exp\left(-\frac{a^2}{2}\right) \cos(\alpha\beta + k(\varphi + \pi/2)), \quad (59)$$

$$a_k^{(-)} = \frac{N_-}{\sqrt{k!}} \exp\left(-\frac{a^2}{2}\right) a^k \left(exp(i\alpha\beta + ik\varphi + ik\pi) - exp(-i\alpha\beta - ik\varphi)\right) =$$
$$\frac{N_-}{\sqrt{k!}} \exp\left(-\frac{a^2}{2}\right) a^k exp(ik\pi/2) \left(exp(i\alpha\beta + ik\varphi + ik\pi/2) - exp(-i\alpha\beta - ik\varphi - ik\pi/2)\right) =$$
$$i \frac{N_+}{\sqrt{k!}} 2(ia)^k \exp\left(-\frac{a^2}{2}\right) \sin(\alpha\beta + k(\varphi + \pi/2)). \quad (60)$$

We neglect the overall phase factor $i$ in (A14) that does not affect anything and get the final expressions for the amplitudes of the SCS in polar coordinates in Eqs. (9, 10).

**Derivation of formula (36) for the case with $m = 1$ and $m = 2$.** First, consider the simplest case with $m = 1$ for which there are two modes: mode 0 and mode 1. Let the states incoming to the beam splitter $BS_{01}$, which has transmission (reflection) coefficient $t_1$ ($r_1$), be $|k_0\rangle_0 |k_1\rangle_1 = |k_0 k_1\rangle_{01}$ with $k_0, k_1 \geq 0$ being photon numbers. The beam splitter acts on creation operators like this

$$a_0^+ \to t_1 a_0^+ + r_1 a_1^+, \quad (61)$$
$$a_1^+ \to -r_1^* a_0^+ + t_1^* a_1^+. \quad (62)$$

By virtue of (61) and (62), after the beam splitter the input states $|k_0 k_1\rangle_{01}$ is transformed to

$$BS_{01} |k_0 k_1\rangle_{01} = \frac{1}{\sqrt{k_0! k_1!}} BS_{01} \left(a_0^{+k_0} a_1^{+k_1}\right) |00\rangle_{01} =$$
$$\frac{(t_1 a_0^+ + r_1 a_1^+)^{k_0} (-r_1^* a_0^+ + t_1^* a_1^+)^{k_1}}{\sqrt{k_0! k_1!}} |00\rangle_{01}. \quad (63)$$

The action of the displacement operator $D_1(\alpha_1)$ on mode 1 of the state (63) can be written as

$$D_1(\alpha_1) BS_{01} |k_0 k_1\rangle_{01} = \frac{1}{\sqrt{k_0! k_1!}}$$

$$D_1(\alpha_1)(t_1 a_0^+ + r_1 a_1^+)^{k_0} D_1^+(\alpha_1) D_1(\alpha_1)(-r_1^* a_0^+ + t_1^* a_1^+)^{k_1} D_1^+(\alpha_1) D_1(\alpha_1) |00\rangle_{01} \quad (64)$$

thanks to the identity $D_1^+(\alpha_1) D_1(\alpha_1) = 1$. Next, using the properties $D_1(\alpha_1) a_1^+ D_1^+(\alpha_1) = a_1^+ - \alpha_1^*$ and $D_1(\alpha_1) |0\rangle_1 = |\alpha_1\rangle_1 = exp\left(-\frac{|\alpha_1|^2}{2}\right) \sum_{l=0}^{\infty} \frac{\alpha_1^l}{\sqrt{l!}} |l\rangle_1$ we bring (64) to (63) to get

$$D_1(\alpha_1) BS_{01} |k_0 k_1\rangle_{01} = \frac{(t_1 a_0^+ + r_1 (a_1^+ - \alpha_1^*))^{k_0} (-r_1^* a_0^+ + t_1^* (a_1^+ - \alpha_1^*))^{k_1}}{\sqrt{k_0! k_1!}} |0\rangle_0$$
$$\cdot exp\left(-\frac{|\alpha_1|^2}{2}\right) \sum_{l=0}^{\infty} \frac{\alpha_1^l}{\sqrt{l!}} |l\rangle_1. \quad (65)$$

We are interested in the situation when neither detectors click (i.e., no photons are registered at all the detectors). In such situation the post-selected state reads (by formally replacing $a_1^+$ by zero in (65))

$$|\Gamma_n^{(1)}\rangle_0 = \frac{1}{\sqrt{P_n^{(1)}}} \frac{(t_1)^{k_0}(-r_1^*)^{k_1}\left(a_0^+ - \frac{r_1}{t_1}\alpha_1^*\right)^{k_0}\left(a_0^+ - \frac{-t_1^*}{r_1^*}\alpha_1^*\right)^{k_1}}{\sqrt{k_0!k_1!}} \exp\left(-\frac{|\alpha_1|^2}{2}\right)|0\rangle_0 =$$

$$\frac{1}{\sqrt{P_n^{(1)}}} \frac{(t_1)^{k_0}(-r_1^*)^{k_1}}{\sqrt{k_0!k_1!}} \left[D_0\left(\frac{r_1^*}{t_1^*}\alpha_1\right)a_0^+ D_0^\dagger\left(\frac{r_1^*}{t_1^*}\alpha_1\right)\right]^{k_0} \left[D_0\left(\frac{-t_1}{r_1}\alpha_1\right)a_0^+ D_0^\dagger\left(\frac{-t_1}{r_1}\alpha_1\right)\right]^{k_1}$$

$$\times \exp\left(-\frac{|\alpha_1|^2}{2}\right)|0\rangle_0, \tag{66}$$

where $n = k_0 + k_1$ and

$$P_n^{(1)} = \frac{1}{k_0!k_1!}\exp(-|\alpha_1|^2)\sum_{k=0}^{n}|m_k(t_1, r_1, \alpha_1)|^2 k!, \tag{67}$$

is the success probability. Here, the amplitudes $m_k(t_1, r_1, \alpha_1)$ are obtained by expanding the expression $(t_1 a_0^+ - r_1 \alpha_1^*)^{k_0}(-r_1^* a_0^+ - t_1^* \alpha_1^*)^{k_1} = \sum_{k=0}^{n} m_k(t_1, r_1, \alpha_1) a_0^{+k}$ in powers of the creation operator $a_0^+$. We do not provide analytical expressions for $m_k(t_1, r_1, \alpha_1)$ because of their complexity of representation. However, these expressions can be directly obtained in numerical simulation. If we define $N_n^{(1)}$ and $\beta_k^{(1)}$ as in (43) and (44), we can rewrite $|\Gamma_n^{(1)}\rangle_0$ in the following form

$$|\Gamma_n^{(1)}\rangle_0 = N_n^{(1)} \prod_{k=1}^{n} D_0(\beta_k^{(1)*}) a^+ D_0^\dagger(\beta_k^{(1)*})|0\rangle_0, \tag{68}$$

which upon action of $D_0(i\alpha)$ on mode 0 yields the output state $|\Omega_n^{(m)}\rangle_0$ of Eq. (36) for $m = 1$.

Now, consider the case of $m = 2$ for which there are three modes: the principal mode 0 and two auxiliary modes 2 and 3. The input state is $|k_0 k_1 k_2\rangle_{012}$, with photon numbers $k_0, k_1, k_2 \geq 0$. Two beam splitters with parameters $(t_1, r_1)$ and $(t_2, r_2)$ are used to mix modes 0, 1 and modes 0, 2, respectively,

$$BS_{01}^{(1)} BS_{02}^{(2)} |k_0 k_1 k_2\rangle_{012} = \frac{BS_{02}^{(2)} BS_{01}^{(1)} a_0^{+k_0} a_1^{+k_1} a_2^{+k_2}}{\sqrt{k_0!k_1!k_2!}}|000\rangle_{012} =$$

$$\frac{(t_1(t_2 a_0^+ + r_2 a_2^+) + r_1 a_1^+)^{k_0}(-r_1^*(t_2 a_0^+ + r_2 a_2^+) + t_1^* a_1^+)^{k_1}(-r_2^* a_0^+ + t_2^* a_2^+)^{k_2}}{\sqrt{k_0!k_1!k_2!}}|000\rangle_{012}. \tag{69}$$

A subsequent unitary operation is associated with two displacement operators $D_1(\alpha_1)$ and $D_2(\alpha_2)$ that transform the state (69) into

$$D_1(\alpha_1) D_2(\alpha_2) BS_{01}^{(1)} BS_{02}^{(2)} |k_0 k_1 k_2\rangle_{012} =$$

$$= \frac{1}{\sqrt{k_0!k_1!k_2!}} \cdot \left(t_1(t_2 a_0^+ + r_2(a_2^+ - \alpha_2^*)) + r_1(a_1^+ - \alpha_1^*)\right)^{k_0},]$$

$$\times \left(-r_1^*(t_2 a_0^+ + r_2(a_2^+ - \alpha_2^*)) + t_1^*(a_1^+ - \alpha_1^*)\right)^{k_1}$$

$$\times \left(-r_2^* a_0^+ + t_2^*(a_2^+ - \alpha_2^*)\right)^{k_2}|0\alpha_1 \alpha_2\rangle_{012}. \tag{70}$$

If we are again interested in generating the conditional state when no clicks are seen in the auxiliary modes (the state (70) is projected onto $|00\rangle_{12}$), then we can formally replace the creation operation $a_2^+$ by zero in formula (B10) to obtain the state

$$|\Gamma_n^{(2)}\rangle_0 = \frac{1}{\sqrt{P_n^{(2)}}} \frac{(t_1 t_2)^{k_0}(-r_1^* t_2)^{k_1}(-r_2^*)^{k_2}}{\sqrt{k_0!k_1!k_2!}} \left(a_0^+ - \frac{t_1 r_2 \alpha_2^* + r_1 \alpha_1^*}{t_1 t_2}\right)^{k_0} \left(a_0^+ - \frac{r_1^* r_2 \alpha_2^* - t_1^* \alpha_1^*}{r_1^* t_2}\right)^{k_1}$$

$$\times \left(a_0^+ - \frac{-t_2^* \alpha_2^*}{r_2^*}\right)^{k_2} \exp\left(-\frac{|\alpha_1|^2 + |\alpha_2|^2}{2}\right)|0\rangle_{012} =$$

$$\frac{1}{\sqrt{P_n^{(2)}}} \frac{(t_1 t_2)^{k_0}(-r_1^* t_2)^{k_1}(-r_2^*)^{k_2}}{\sqrt{k_0! k_1! k_2!}} \left[ D_0\left(\frac{t_1^* r_2^* \alpha_2 + r_1^* \alpha_1}{t_1^* t_2^*}\right) a_0^+ D_0^\dagger\left(\frac{t_1^* r_2^* \alpha_2 + r_1^* \alpha_1}{t_1^* t_2^*}\right) \right]^{k_0}$$

$$\left[ D_0\left(\frac{r_1 r_2^* \alpha_2 - t_1 \alpha_1}{r_1 t_2^*}\right) a_0^+ D_0^\dagger\left(\frac{r_1 r_2^* \alpha_2 - t_1 \alpha_1}{r_1 t_2^*}\right) \right]^{k_1} \left[ D_0\left(-\frac{t_2 \alpha_2}{r_2}\right) a_0^+ D_0^\dagger\left(-\frac{t_2 \alpha_2}{r_2}\right) \right]^{k_2},$$

$$exp\left(-\frac{|\alpha_1|^2 + |\alpha_2|^2}{2}\right) |0\rangle_0, \quad (71)$$

where $n = k_0 + k_1 + k_2$ and

$$P_n^{(2)} = \frac{1}{k_0! k_1! k_2!} exp(-(|\alpha_1|^2 + |\alpha_2|^2)) \sum_{k=0}^n |m_k(t_1, r_1, t_2, r_2, \alpha_1, \alpha_2)|^2 k!, \quad (72)$$

being the success probability. Here, the amplitudes $m_k(t_1, r_1, t_2, r_2, \alpha_1, \alpha_2)$ follow from the decomposition of operator expression $(t_1(t_2 a_0^+ - r_2 \alpha_2^*) - r_1 \alpha_1^*)^{k_0}(-r_1^*(t_2 a_0^+ - r_2 \alpha_2^*) - t_1^* \alpha_1^*)^{k_1}(-r_2^* a_0^+ + t_2^* \alpha_2^*)^{k_2} = \sum_{k=0}^n m_k(t_1, r_1, \alpha_1) a_0^{+k}$ and are not presented due to their complexity. If we define $N_n^{(2)}$ and $\beta_k^{(2)}$ as in (45) and (46), then we can cast $\left|\Gamma_n^{(2)}\right\rangle_0$ into the following form

$$\left|\Gamma_n^{(2)}\right\rangle_0 = N_n^{(2)} \prod_{k=1}^n D_0\left(\beta_k^{(2)*}\right) a^+ D_0^\dagger\left(\beta_k^{(2)*}\right) |0\rangle_0, \quad (73)$$

which, upon the action of $D_0(i\alpha)$ on the principal mode 0, is nothing else but the output state $\left|\Omega_n^{(m)}\right\rangle_0$ of Eq. (36) for $m = 2$.

Likewise, the formula (36) can be derived analytically for any $m > 2$. However, the formulation gets more cumbersome and thus will not be presented explicitly.

**References**

1. P. Shor, Proceedings of the 35th Annual Symposium on Foundation of Computer Science. IEEE, Computer Society Press, Santa Fe, NM, (1994).
2. L. K. Grover, "Quantum mechanics helps in searching for a needle in a haystack", Phys. Rev. Lett. **79**, 325-328 (1997).
3. A. Barenco, C. Bennett, R. Cleve, D. P. DiVincenzo, N. Margolus, P. Shor, T. Sleator, J. Smolin, and H. Weinfurter, "Elementary gates for quantum computation", Phys. Rev. A **52**, 3457-3467 (1995).
4. D. Gottesman and I. L. Chuang, "Demonstrating the viability of universal quantum computation using teleportation and single-qubit operations", Nature **402**, 390-393 (1999).
5. E. Knill, L. Laflamme and G. J. Milburn, "A scheme for efficient quantum computation with linear optics", Nature **409**, 46-52 (2001).
6. R. Rausendorf and H. J. Briegel, "A one-way quantum computer", Phys. Rev. Lett. **86**, 5188-5191 (2001).
7. J. L. O'Brien J L, A. Furusawa and J. Vučković, "Photonic quantum technologies", Nat. Photon. **3** 687-695 (2009).
8. S. Braunstein and P. van Loock, "Quantum information with continuous variables", Rev. Mod. Phys. **77,** 513-577 (2005).
9. A. Furusava, P. van Loock, "Quantum teleportation and experiment-A hybrid approach to optical quantum information processing", Wiley-VCH, Weinheim (2011).
10. Z. X. Man, Y. J. Xia, N. B. An, "Simultaneous observation of particle and wave behaviors of entangled photons", Scientific Repots **7**, 42539 (2017).
11. A. S. Rab, E. Polino, Z. X. Man, N. B. An, Y. J. Xia, N. Spagnolo, R. L. Franco, F. Sciarrino, "Entanglement of photons in their dual wave-particle nature", Nature Communications **8**, 915 (2017).

12. N. Lutkenhaus, J. Calsamiglia, K. A. Suominen, "Bell measurements for teleportation", Phys. Rev. A **59**, 3245 (1999).
13. S. L. Braunstein, H. J. Kimble, "Teleportation of continuous quantum variables", Phys. Rev. Lett. **80**, 869 (1998).
14. S. J. van Enk and O. Hirota, "Entangled coherent states: teleportation and decoherence", Phys. Rev. A **64**, 022313 (2001).
15. N. B. An, "Teleportation of coherent state superposition within a network", Phys. Rev. A **68**, 022321 (2003).
16. N. B. An, "Optimal processing of quantum information via W-type entangled coherent states", Phys. Rev. A **69**, 022315 (2004).
17. S. A. Podoshvedov, "Generation of displaced squeezed superpositions of coherent states", J. Exp. Theor. Phys. **114**, 451-464 (2012).
18. S. A. Podoshvedov, "Displaced rotations of coherent states", Quant. Inf. Proc. **11**, 1809-1828 (2012).
19. H. N. Phien and N. B. An, "Quantum teleportation of an arbitrary two-mode coherent state using only linear optics elements", Phys. Lett. A **372**, 2825-2829 (2008).
20. N. B. An, "Teleportation of a general two-mode coherent-state superposition via attenuated quantum channels with ideal and/or threshold detectors", Phys. Lett. A **373**, 1701 (2009).
21. S. A. Podoshvedov, J. Kim, K. Kim, "Elementary quantum gates with Gaussian states", Quan. Inf. Proc. **13**, 1723-1749 (2014).
22. Nguyen Ba An and Jaewan Kim, "Cluster-type entangled coherent states: Generation and application", Phys. Rev. A **80**, 042316 (2009).
23. N. B. An, K. Kim and J. Kim, "Generation of cluster-type entangled coherent states using weak nonlinearities and intense laser beams", Quantum Inf. Comput. **11**, 0124 (2011).
24. S. A. Podoshvedov, "Single qubit operations with base squeezed coherent states", Optics Commun. **290**, 192-201 (2013).
25. S. A. Podoshvedov, "Building of one-way Hadamard gate for squeezed coherent states", Phys. Rev. A **87**, 012307 (2013).
26. K. Huang, H. Le Jeannic, J. Ruaudel, V. B. Verms, M. D. Shaw, F. Marsili, S. W. Nam, E, Wu, H. Zeng, Y.-C. Jeong, R. Filip, O. Moring and J. Laurat, "Optical synthesis of large-amplitude squeezed coherent-state superpositions with minimal resources", Phys. Rev. Lett. **115**, 023602 (2015).
27. N. Lee, H. Benichi, Y. Takeno, S. Takeda, J. Webb, E. Huntington, A. Furusawa, "Teleportation of nonclassical wave packets of light" Science **352**, 330-333 (2011).
28. S. Takeda, T. Mizuta, M. Fuwa, P. van Loock, A. Furusawa, "Deterministic quantum teleportation of photonic quantum bits by a hybrid technique", Nature **500**, 315-318 (2013).
29. S. A. Podoshvedov, "Quantum teleportation protocol with an assistant who prepares amplitude modulated unknown qubit", JOSA B **35**, 861-877 (2018).
30. S. A. Podoshvedov, Nguyen Ba An, "Designs of interactions between discrete- and continuous-variable states for generation of hybrid entanglement", Quantum Inf. Process. **18**, 68 (2019).
31. A. I. Lvovsky, R. Ghobadi, A. Chandra, A. S. Prasad, C. Simon, "Observation of micro-macro entanglement of light", Nature Phys. **9**, 541-544 (2013).
32. P. Sekatski, N. Sangouard, M. Stobinska, F. Bussieres, M. Afzelius, N. Gisin, "Proposal for exploring macroscopic entanglement with a single photon and coherent states", Phys. Rev. A **86**, 060301 (2012).
33. O. Morin, K. Haung, J. Liu, H. L. Jeannic, C. Fabre, J. Laurat, "Remote creation of hybrid entanglement between particle-like and wave-like optical qubits", Nature Photonics **8**, 570-574 (2014).

34. H. Le Jeannic, A. Cavailles, J. Raskop, K. Huang, J. Laurat, "Remote preparation of continuous-variable qubits using loss-tolerant hybrid entanglement of light", Optica **5**, 1012-1015 (2018).
35. S. Bruno, A. Martin, P. Sekatski, N. Sangouard, R. Thew, N. Gisin, "Displacement of entanglement back and forth between the micro and macro domains", Nature Phys. **9**, 545-550 2013.
36. S. A. Podoshvedov, "Efficient quantum teleportation of unknown qubit based on DV-CV interaction mechanism", Entropy **21**, 150 (2019).
37. B. C. Sanders, "Entangled coherent states", Phys. Rev. A **45**, 6811-6815 (1992).
38. B. Yurke and D. Stoler, "Generating quantum mechanical superpositions of macroscopically distinguishable states via amplitude dispersion", Phys. Rev. Lett. **57**, 13-17 (1986).
39. M. Dakna, T. Anhut, T. Opatrn´y, L. Knöll, and D.-G. Welsch, "Generating Schrödinger-cat-like states by means of conditional measurements on a beam splitter", Phys. Rev. A **55**, 3184-3194 (1997).
40. M. Dakna, J. Clausen, L. Knöll, and D.-G. Welsch, "Generation of arbitrary quantum states of traveling fields", Phys. Rev. A **59**, 1658-1661 (1999).
41. A. P. Lund, H. Jeong, T. C. Ralph and M. S. Kim, "Conditional production of superpositions of coherent states with inefficient photon detection", Phys. Rev. A **70**, 020101 (2004).
42. P. Marek, H. Jeong and M. S. Kim, "Generating "squeezed" superpositions of coherent states using photon addition and subtraction", Phys. Rev. A **78**, 063811 (2008).
43. S. A. Podoshvedov, "Schemes for performance of displacing Hadamard gate with coherent states", Optics Commun. **285**, 3896–3905 (2012).
44. J. Wenger, R. Tualle-Brouri and P. Grangier, "Non-Gaussian statistics from individual pulses of squeezed vacuum", Phys. Rev. Lett. **92**, 153601 (2004).
45. A. Ourjoumtsev, R. Tualle-Brouri, J. Laurat and P. Grangier, "Generating optical Schrödinger kittens for quantum information processing", Science **312**, 83-86 (2006).
46. J. S. Neergaard-Nielsen, M. Nielsen, C. Hettich, K. Mølmer, and E. S. Polzik, "Generation of a superposition of odd photon number states for quantum information networks", Phys. Rev. Lett. **97**, 083604 (2006).
47. A. Ourjoumtsev, F. Ferreyrol, R. Tualle-Brouri and P. Grangier, "Preparation of non-local superpositions of quasi-classical light states", Nat. Phys. **5**, 189-192 (2009).
48. A. Tipsmark, R. Dong, A. Laghaout, P. Marek, M. Jezek, and U. L. Andersen, "Experimental demonstration of a Hadamard gate for coherent state qubits," Phys. Rev. A **84**, 050301(R) (2011).
49. T. Gerrits, S. Glancy, T. S. Clement, B. Calkins, A. E. Lita, A. J. Miller, A. L. Migdall, S. W. Nam, R. P. Mirin, and E. Knill, "Generation of optical coherent-state superpositions by number-resolved subtraction from the squeezed vacuum," Phys. Rev. A **82**, 031802 (2010).
50. J.-I. Yoshikawa, M. Bergmann, P. van Loock, M. Fuwa, M. Okada, K. Takase, T. Toyama, K. Makino, S. Takeds and A. Furusawa, "Heralded creation of photonic qudits from parametric down conversion using linear optics", Phys Rev. A **97**, 053814 (2018).
51. S. A. Podoshvedov, "Extraction of displaced number states", JOSA B **31**, 2491-2503 (2014).
52. S. A. Podoshvedov, "Elementary quantum gates in different bases", Quant. Inf. Proc. **15**, 3967-3993 (2016).
53. M. D. A. Paris, "Displacement operator by beam splitter", Phys. Lett. A **217**, 78-81 (1996).
54. A. I. Lvovsky and S. A. Babichev, "Synthesis and tomographic characterization of the displaced Fock state of light", Phys. Rev. A **66**, 011801(R) 2002.

55. Knill, E; Laflamme, L; Milburn. G.J. A scheme for efficient quantum computation with linear optics. Nature **2001**, *409*, 46-52.
56. D. F. Walls and G. J. Milburn, Quantum Optics, Springer-Verlag, Berlin Heidelberg (1994).

## Acknowledgement

S.A.P. is supported by Act 211 Government of the Russian Federation, contract № 02.A03.21.0011, while N.B.A. is supported by the National Foundation for Science and Technology Development (NAFOSTED) under project no. 103.01-2017.08.

## Author Contributions

S.A.P developed the theory of $\alpha-$representation of SCS and proposed the concept of SCQ. S.A.P. and N.B.A. equally contributed to development of mathematical apparatus and accuracy of mathematical calculations. S.A.P. proposed idea of SCQ generation with use of two-mode entangled state (17). E.V.M. discovered idea of SCQ generation from Fock states and developed in numerically with help of S.A.P. and N.B.A. E.V.M., A.S.P and D.A.K. executed all numerical simulations. Both (S.A.P. and N.B.A.) wrote and reviewed the manuscript.

## Additional Information

**Competing financial and/or non-financial interests:** We declare that the authors have no competing (financial) interests as defined by Nature Research, or other (non-financial) interests that might be perceived to influence the results and/or discussion reported in this paper.

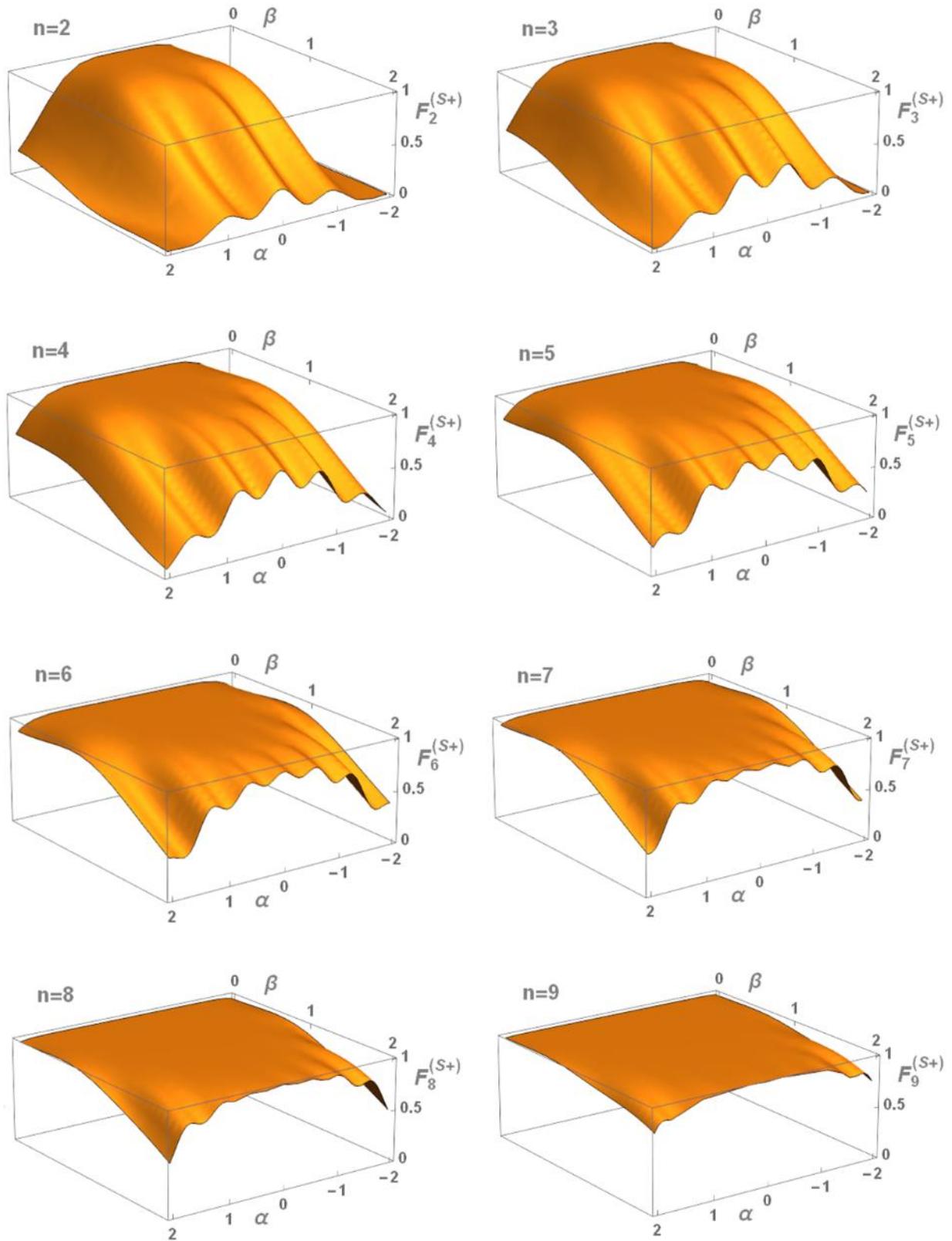

**Fig. 1.** Fidelity $F_n^{(S+)}$ (15) between even SCS (1) and its truncated version (11) in dependency on its size $\beta$ and displacement amplitude $\alpha$ of the base elements. From top to bottom and from left to right, the SCS dimension grows from $n = 2$ up to $n = 9$.

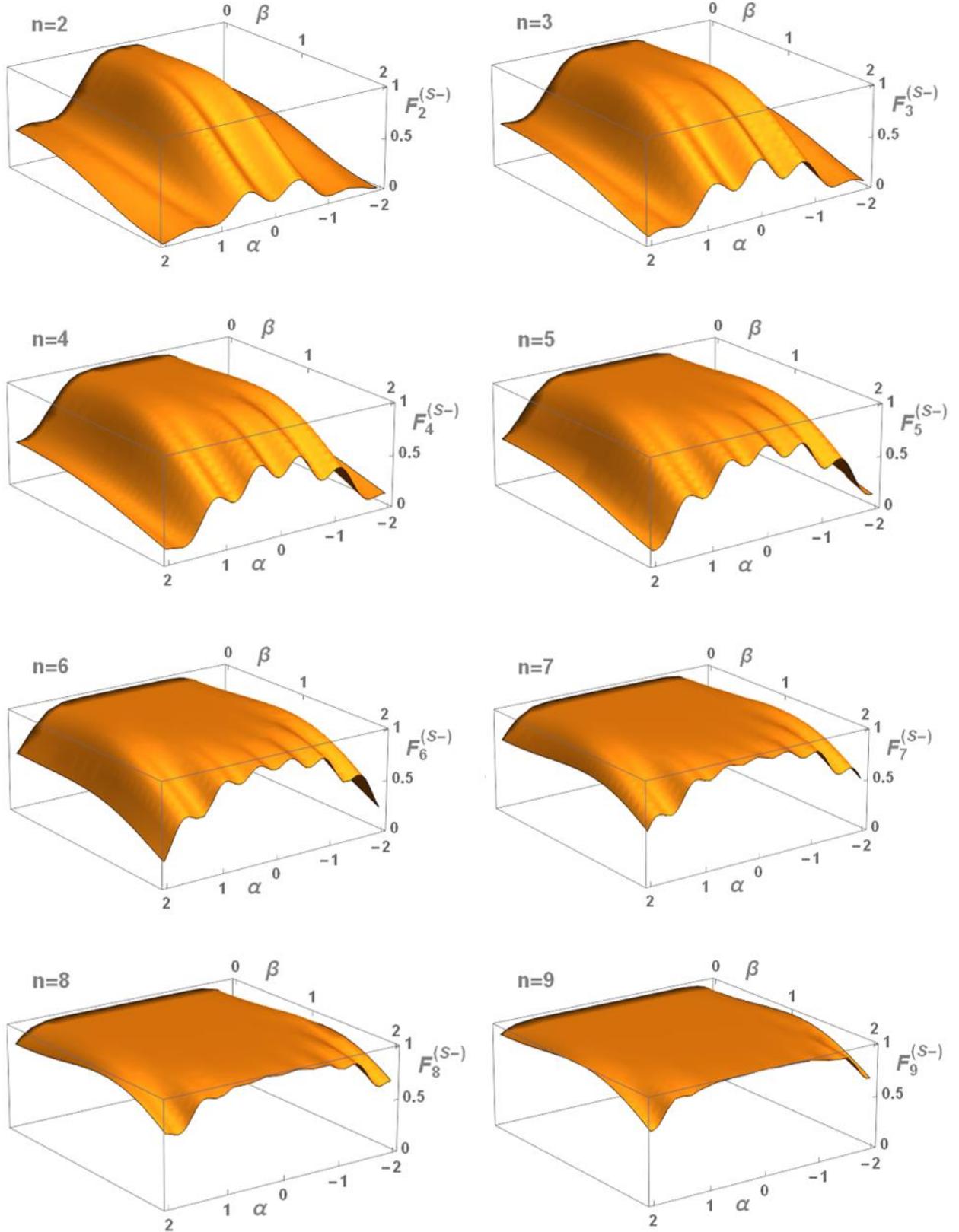

**Fig. 2.** Fidelity $F_n^{(S-)}$ (15) between even SCS (2) and its truncated version (11) in dependency on its size $\beta$ and displacement amplitude $\alpha$ of the base elements. From top to bottom and from left to right, the SCS dimension grows from $n = 2$ up to $n = 9$.

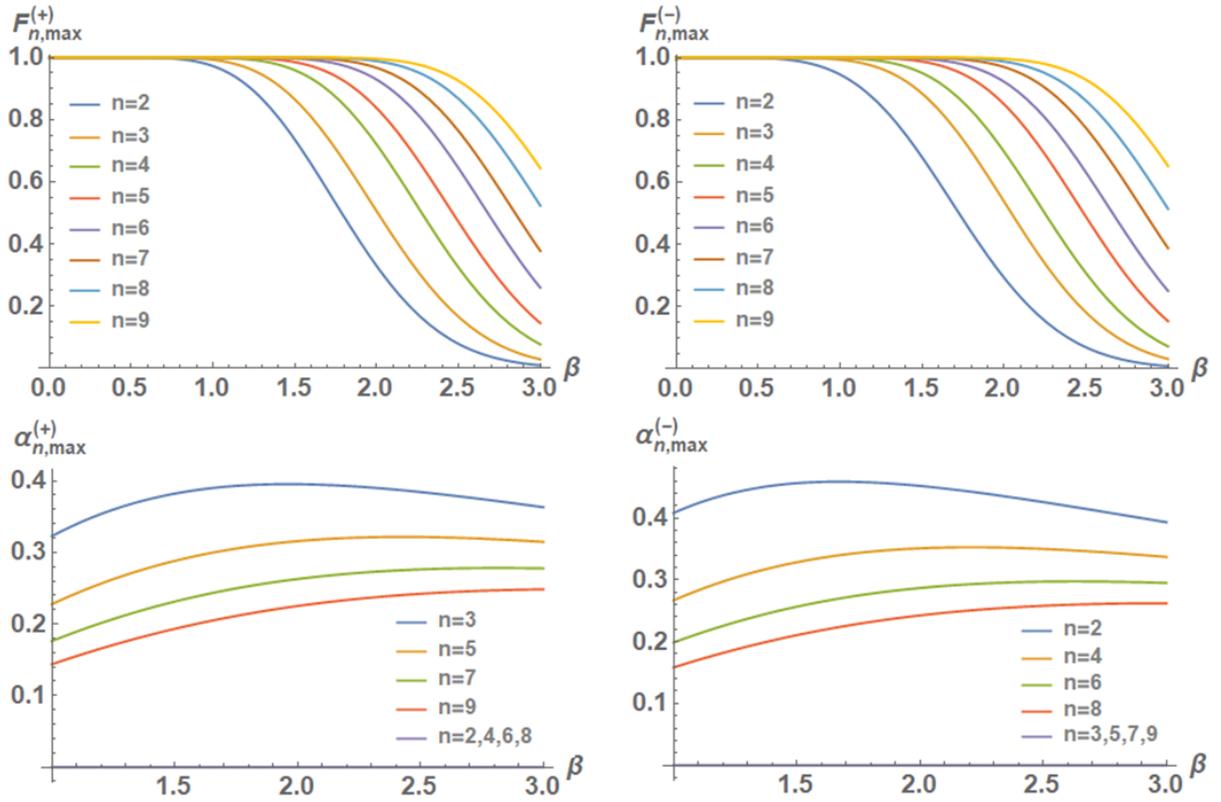

**Fig. 3.** Maximal fidelities $F^{(S+)}_{n,max}$ (top-left) and $F^{(S-)}_{n,max}$ (top-right) (Eq. (15)) between SCSs (1, 2) and SCQ (11, 12) against its size $\beta$. The displacement amplitude $\alpha^{(+)}_{n,max}$ (bottom-left) and $\alpha^{(-)}_{n,max}$ (bottom-right) of the base elements under which the maximal fidelities are observed are shown in dependency on $\beta$.

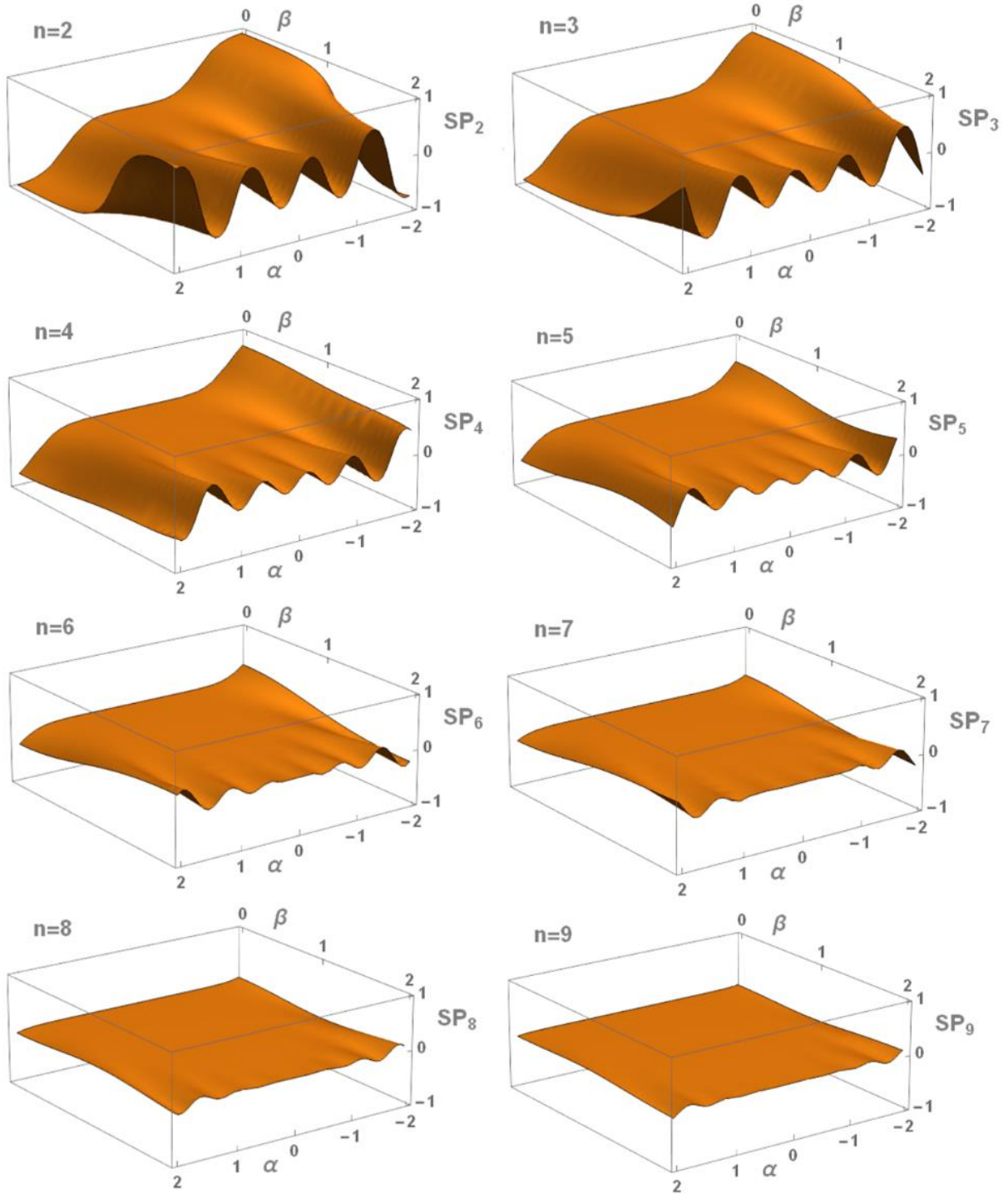

**Fig. 4.** Absolute value of scalar product $SP_n$ (Eq. (16)) between even and odd truncated versions of Schrödinger cats (11, 12) in dependency on its size $\beta$ and displacement amplitude $\alpha$ of the base elements. From top to bottom and from left to right, the scalar product $|SP_n|$ becomes smaller approaching to zero when $n$ grows from $n = 2$ up to $n = 9$.

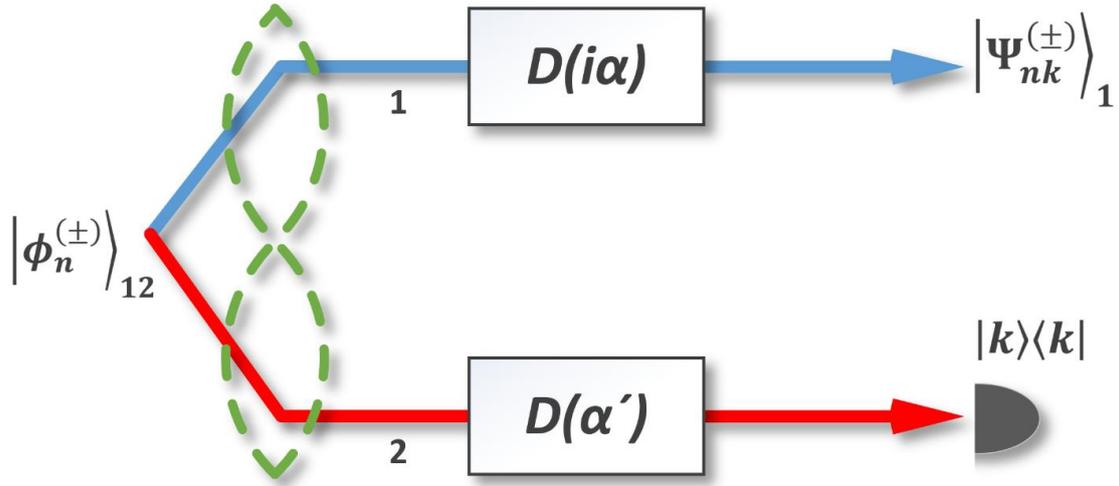

**Fig. 5.** Schematic representation for generation of the states (6), involving SCQs (11, 12). Two HTBS are used to displace initially prepared entangled two-mode states $\left|\phi_n^{(\pm)}\right\rangle_{12}$ of Eq. 17) by quantities $\alpha$ and $\alpha'$, respectively. Conditioned on registration of $k$ photons in mode 2, the initial state $\left|\phi_n^{(\pm)}\right\rangle_{12}$ is projected onto $\left|\Psi_{nk}^{(\pm)}\right\rangle_1$ which may approximate even/odd optical SCSs.

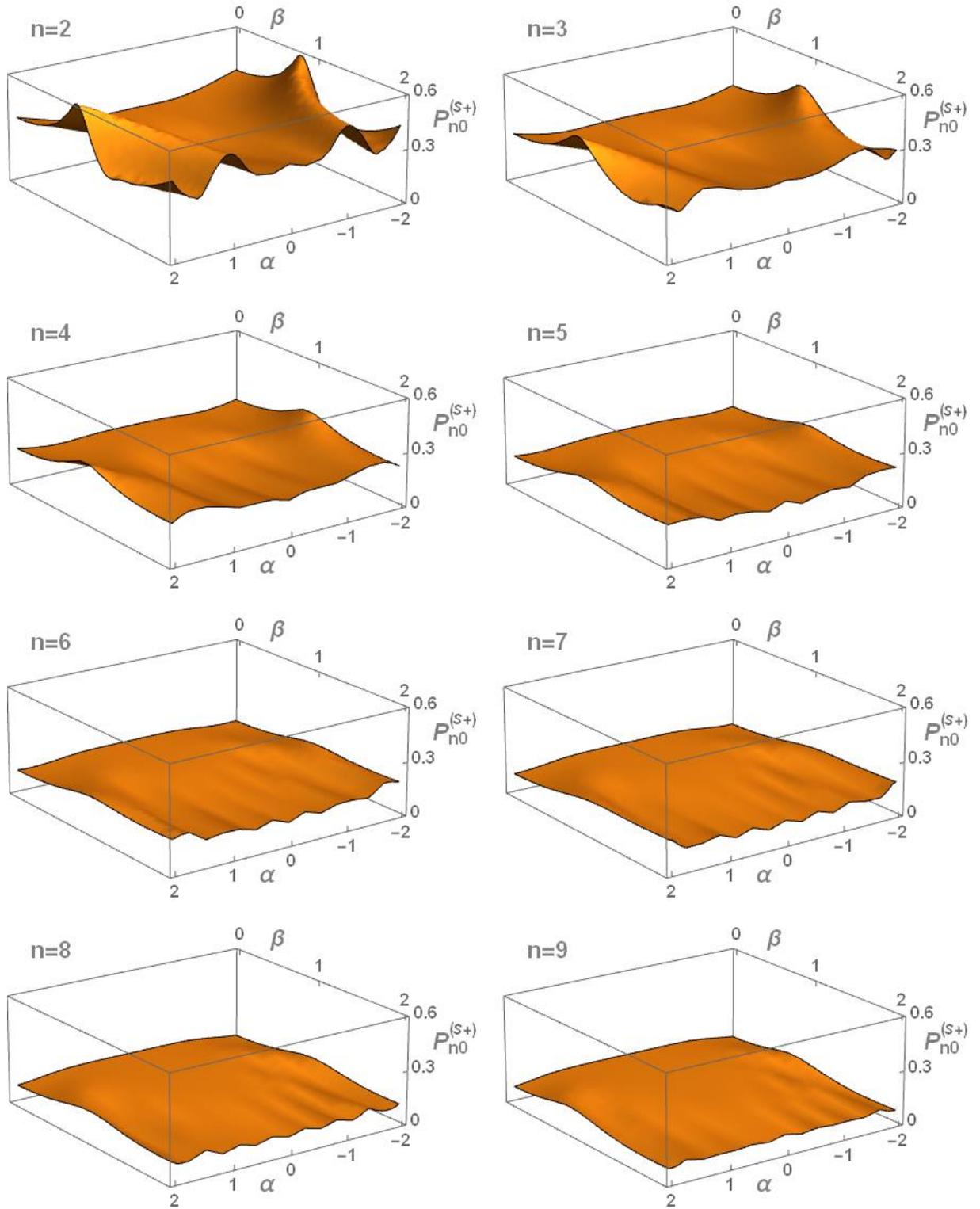

**Fig. 6.** Three-dimensional plots of maximal success probabilities $P_{n0}^{(S+)}$ (Eq. (27)) to generate SCQs (11, 12) in dependency on its size $\beta$ and displacement amplitude $\alpha$. From top to bottom and from left to right, the SCS dimension grows from $n=2$ up to $n=9$.

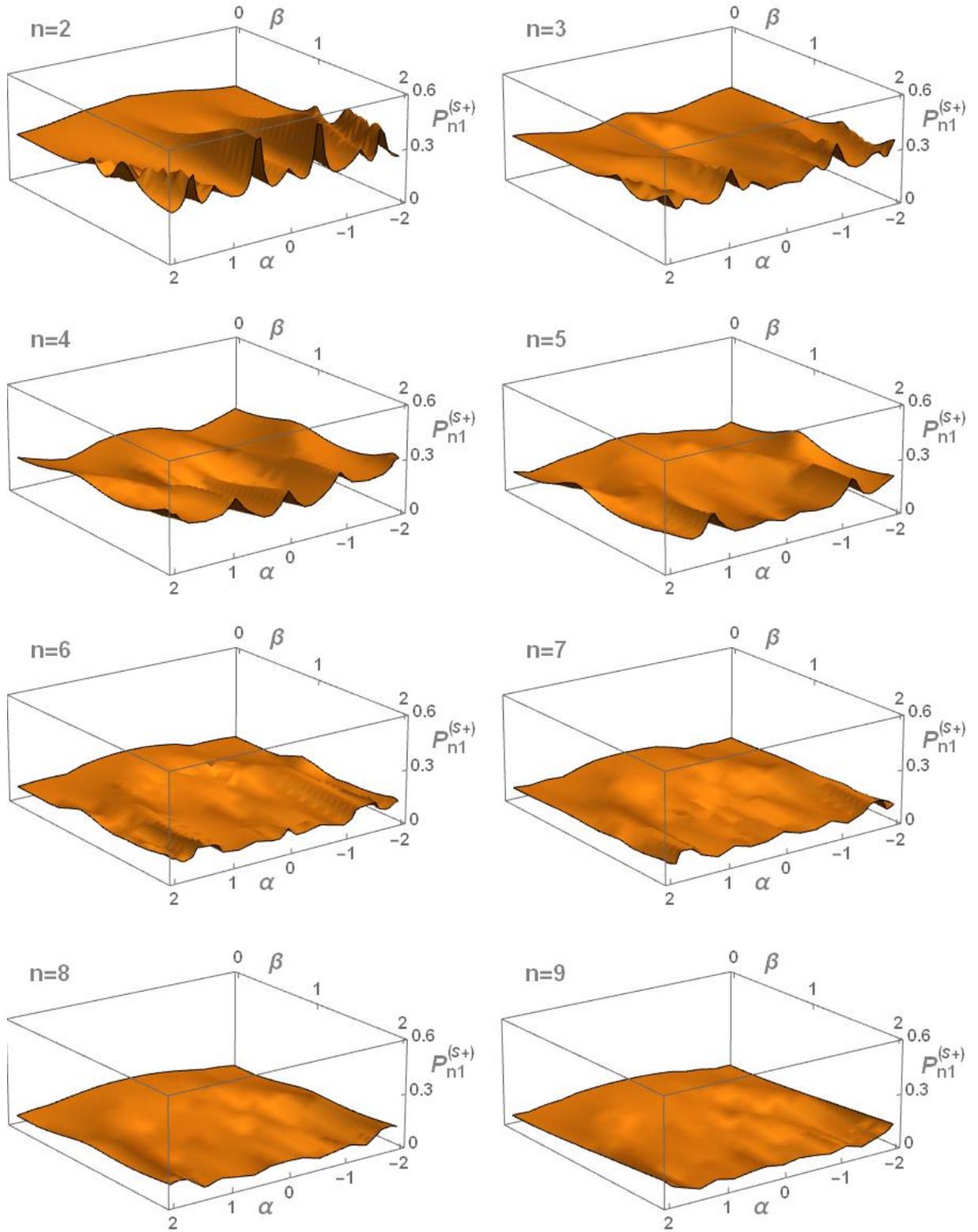

**Fig. 7.** Three-dimensional plots of maximal success probabilities $P_{n1}^{(S+)}$ (Eq. (27)) to generate SCQs (11, 12) in dependency on its size $\beta$ and displacement amplitude $\alpha$. From top to bottom and from left to right, the SCS dimension grows from $n = 2$ up to $n = 9$.

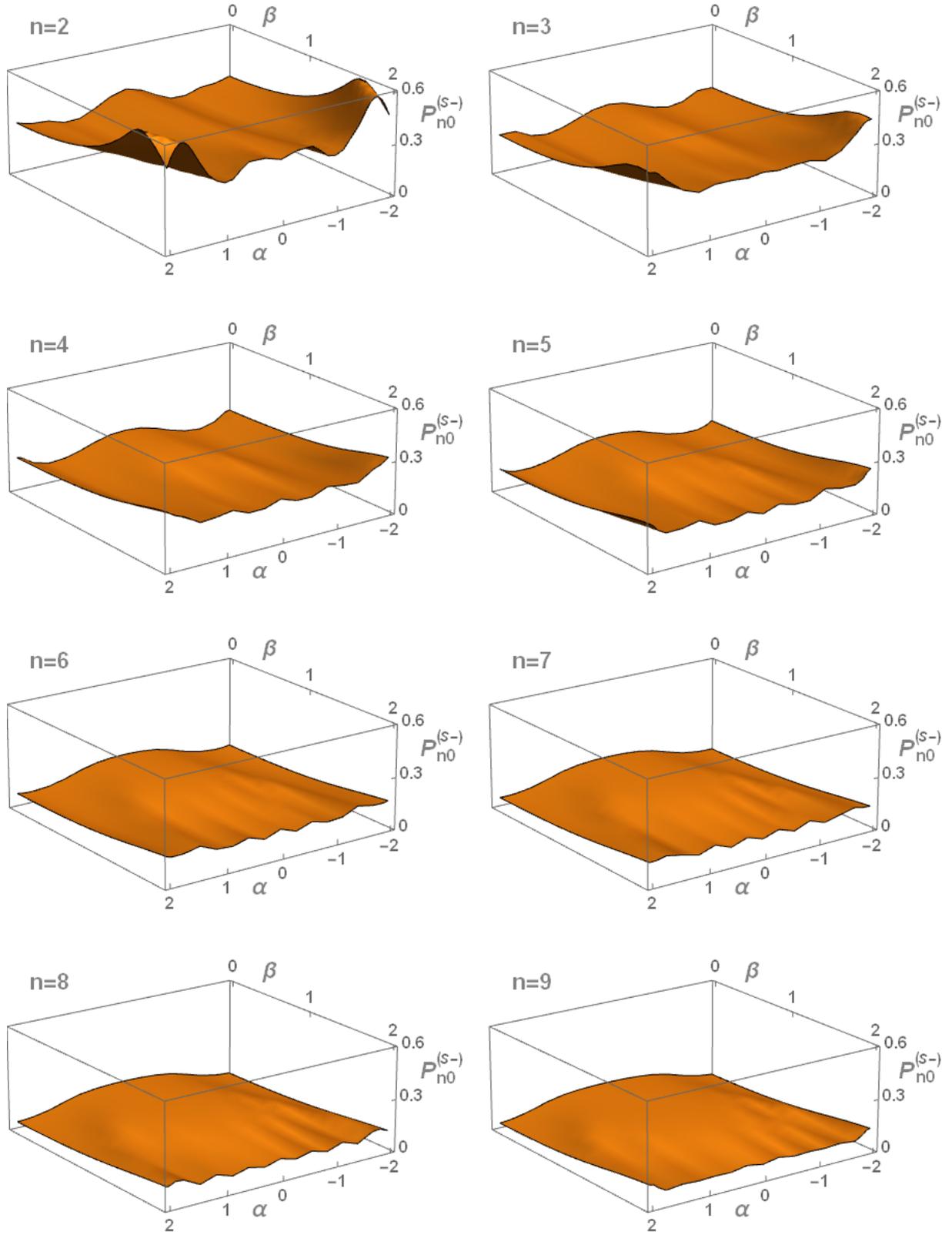

**Fig. 8.** Three-dimensional plots of maximal success probabilities $P_{n0}^{(S-)}$ (Eq. (27)) to generate SCQs (11, 12) in dependency on its size $\beta$ and displacement amplitude $\alpha$. From top to bottom and from left to right, the SCS dimension grows from $n = 2$ up to $n = 9$.

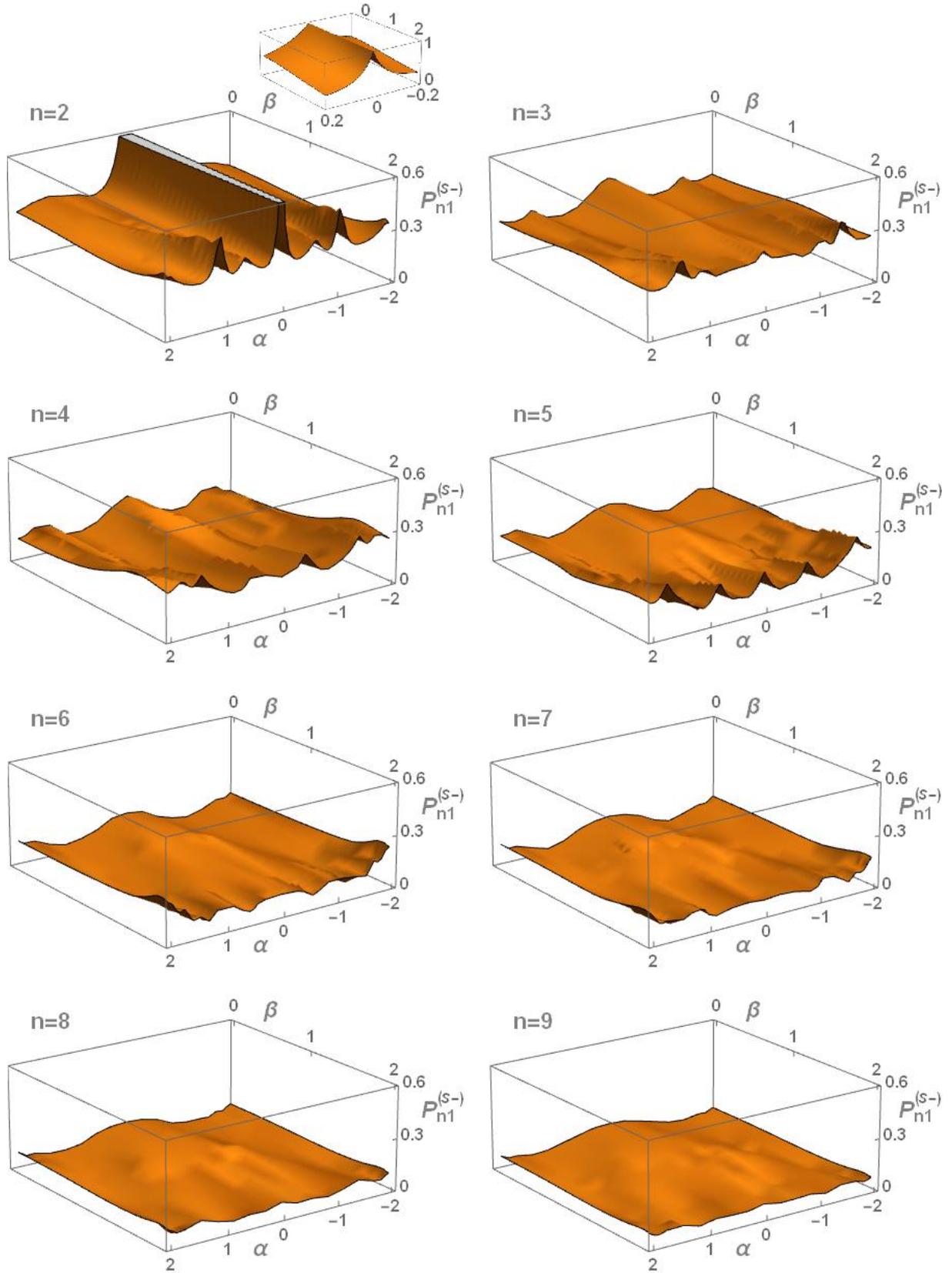

**Fig. 9.** Three-dimensional plots of maximal success probabilities $P_{n1}^{(s-)}$ (Eq. (27)) to generate SCQs (11, 12) in dependency on its size $\beta$ and displacement amplitude $\alpha$. From top to bottom and from left to right, the SCS dimension grows from $n = 2$ up to $n = 9$.

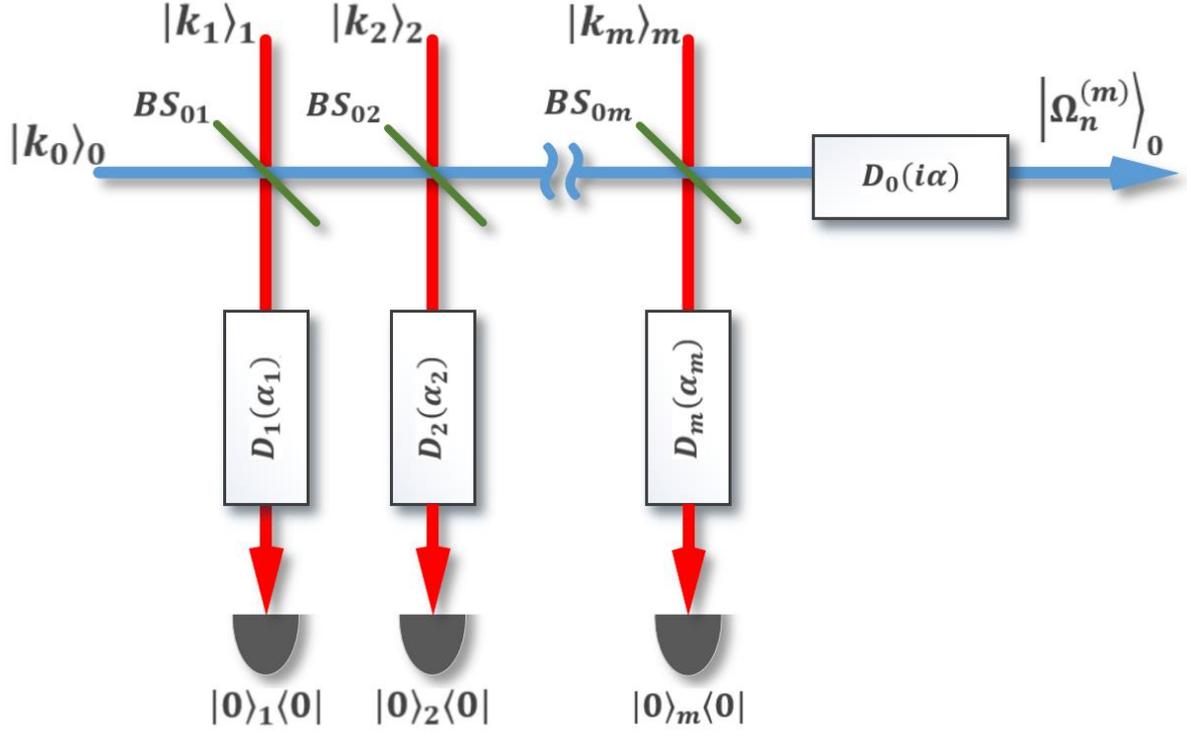

**Fig. 10.** Schematic setup for generation of state $\left|\Omega_n^{(m)}\right\rangle$ in Eq. (36) starting from Fock states (see Section 4). $|k_j\rangle_j$ denotes an input Fock state containing $k_j$ photons in mode $j$, $BS_{0j}$ beam splitter with transmission (reflection) coefficient $t_j$ ($r_j$) acting on mode 0 and mode $j$, $D_j(\alpha)$ displacement operator with displace amplitude $\alpha$ acting on mode $j$, $|0\rangle_j\langle 0|$ implies detection of no photons in mode $j$, and $\left|\Omega_n^{(m)}\right\rangle$ the output post-selected state.

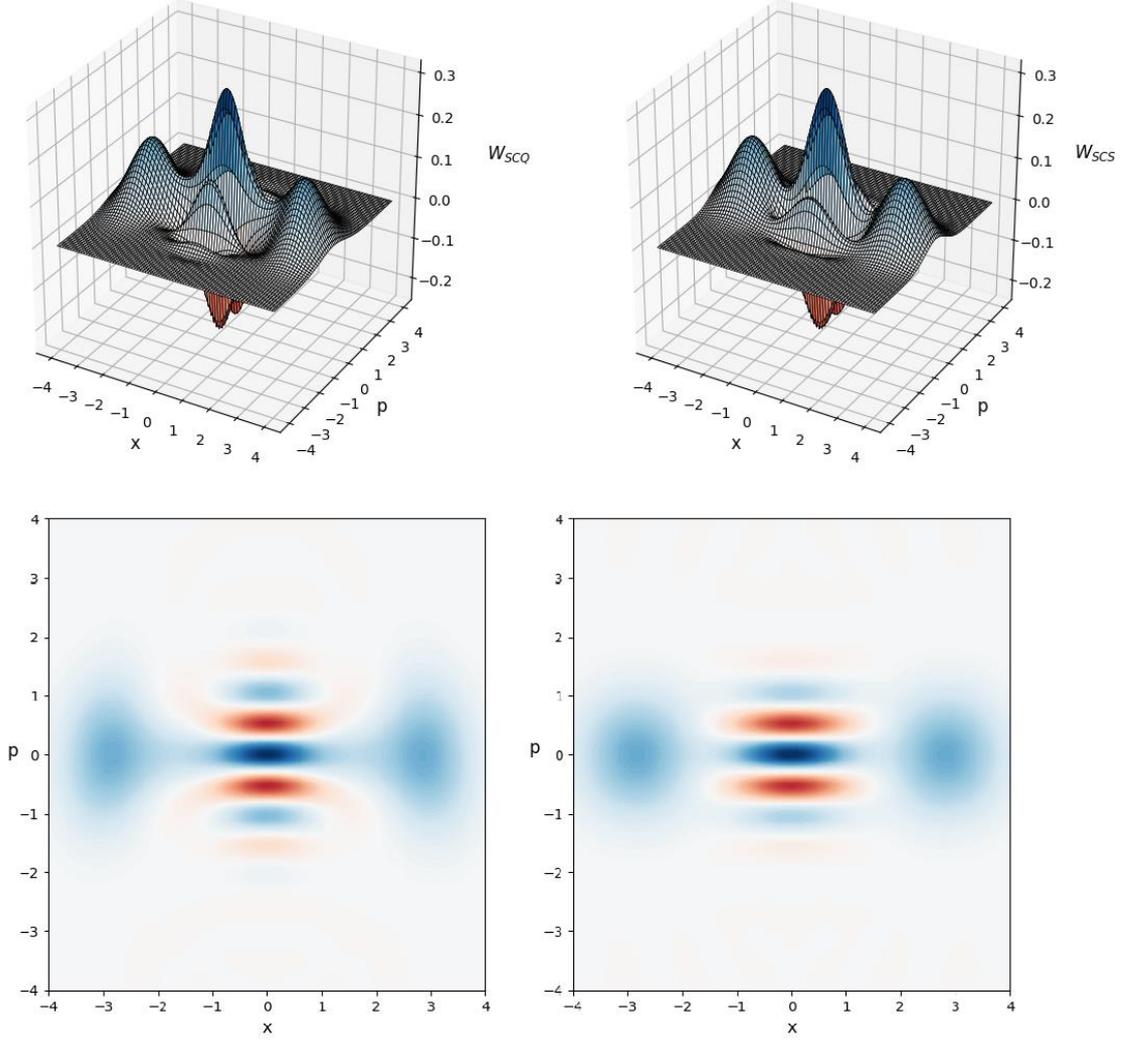

**Fig. 11.** Plot of even Wigner function $W_{SCQ}$ with $m = 3$, $k_0 = 4, k_1 = k_2 = k_3 = 2$ (left-upper subfigure) and its contour image (left-bottom subfigure) generated in optical scheme in Fig. 10 with parameters taken from Table 4 are compared with genuine even Wigner function $W_{SCS}$ with size $\beta = 2$ (right-upper subfigure) and its contour image (right-bottom subfigure). The fidelity calculated by using Wigner functions of generated and genuine states gives the result $F_{10} = 0.980140336082$ comparable to that presented in Table 4.

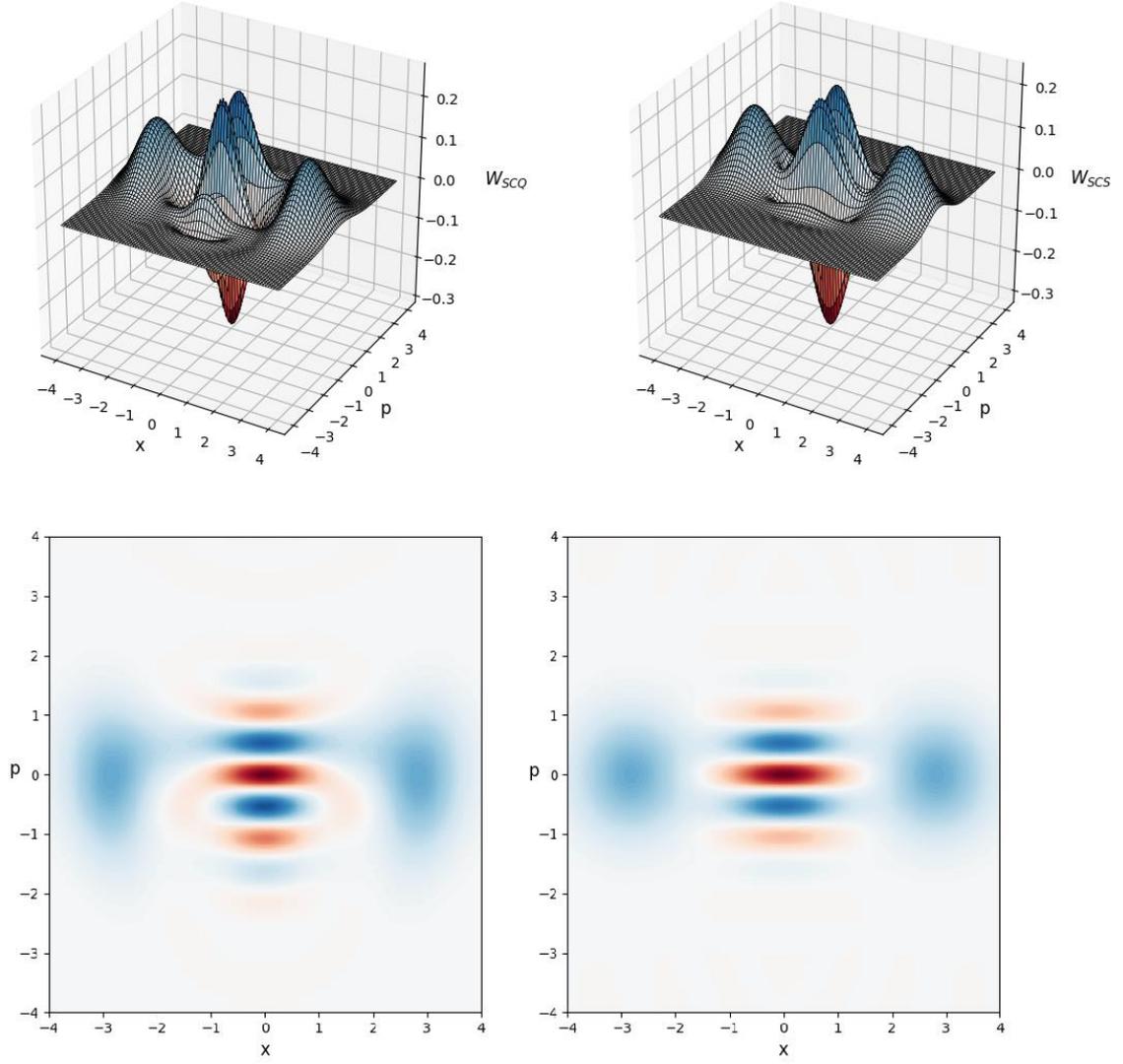

**Fig. 12.** Plot of odd Wigner function $W_{SCQ}$ with $m = 3$, $k_0 = 4, k_1 = k_2 = k_3 = 2$ (left-upper subfigure) and its contour image (left-bottom subfigure) generated in optical scheme in Fig. 10 with parameters taken from Table 4 are compared with genuine odd Wigner function $W_{SCS}$ with size $\beta = 2$ (right-upper subfigure) and its contour image (right-bottom subfigure). The fidelity calculated by using Wigner functions of generated and genuine states gives the result $F_{10} = 0.961285449744$ comparable to that presented in Table 4.

34